\documentclass[aps,prstab,twocolumn,groupedaddress,nofootinbib,floatfix]
{revtex4-1}

\usepackage{amsmath}
\usepackage{graphicx}
\usepackage[caption=false]{subfig}
\usepackage{color,soul}

\usepackage{relsize}
\usepackage{xspace}

\newcommand{\unit}[1]{\ensuremath{\,\mathrm{#1}}}

\DeclareMathOperator*{\SumInt}{%
\mathchoice%
  {\ooalign{$\displaystyle\sum$\cr\hidewidth$\displaystyle\int$\hidewidth\cr}}
  {\ooalign{\raisebox{.14\height}{\scalebox{.7}{$\textstyle\sum$}}\cr\hidewidth$\textstyle\int$\hidewidth\cr}}
  {\ooalign{\raisebox{.2\height}{\scalebox{.6}{$\scriptstyle\sum$}}\cr$\scriptstyle\int$\cr}}
  {\ooalign{\raisebox{.2\height}{\scalebox{.6}{$\scriptstyle\sum$}}\cr$\scriptstyle\int$\cr}}
}

\begin{document}


\title{
Lattice design and experimental studies of nonlinear resonance at the Cornell Electron Storage Ring
}



\author{S. T. Wang}
\email[]{sw565@cornell.edu}
\author{V. Khachatryan}
\affiliation{Cornell Laboratory for Accelerator-based Sciences and 
  Education, Cornell University, Ithaca, NY 14853}
\author{P. Nishikawa}
\affiliation{High Energy Accelerator Research Organization (KEK), Oho, Tsukuba-shi, Ibaraki-ken 305-0801, Japan}




\begin{abstract}
The predominant source of nonlinearity in most existing accelerators are sextupoles, which introduce nonlinear resonances. When the horizontal tune of an accelerator is near such a resonance line ($n\nu_x$), stable fixed points (SFPs) may appear in the horizontal phase space to form a second closed orbit different from the ``zero" closed orbit. The stable islands in phase space surrounding the SFPs are also referred as transverse resonance island ``buckets" (TRIBs). However, the TRIBs are not always present near a resonance line in a storage ring. Necessary conditions for TRIBs formation will be discussed in this paper. A 6-GeV lattice with the horizontal tune near a 3rd-order resonance at $3\nu_{x}=50$ is then designed for the Cornell Electron Storage Ring (CESR). After loading this lattice into CESR, the TRIBs are observed while adjusting the horizontal tune near the 3rd-order resonance, consistent with particle tracking simulations. TRIBs rotation in the horizontal phase space are also experimentally demonstrated with adjustment of a sextupole knob. These results are in good agreement with the theoretical calculation. 
\end{abstract}


\maketitle

\section{Introduction}
Nonlinear resonances are critical topics in the beam dynamics of accelerator physics, which have been studied both theoretically and experimentally for many years \cite{ef_book:1997, sylee_book}. Normally a storage ring operates at tunes far away from resonance lines to avoid negative impacts such as instability, poor lifetime, and increased emittance. However, by taking advantage of the resonance nature, one application of the ring operating near the 4th-order resonance line is to extract particles in multi-turns as at the CERN-PS \cite{cern:2002, cern:2016}. Recently, MLS and BESSY-II have demonstrated the stable two-orbit operation by utilizing the transverse resonance island buckets (TRIBs) for x-ray users \cite{bessy:2015, bessy:2019}. In the second closed orbit depending on the horizontal tune near a 3rd-order or 4th-order resonance line, the beam will travel 3 or 4 turns before returning to its original position in phase space. Thus, the frequency of x-ray pulses from the particle beam populating in one island in the second orbit decreases by 3 or 4 times which expands the opportunities for timing experiments \cite{bessy:2019}. In addition, the spatial separation of two island orbits plus the usage of a twin elliptical undulator provide a unique method of flipping the helicity of x-rays at very high rates (2~\unit{ns}) \cite{hel_switch:2020}.  

TRIBs form in the vicinity of a resonance line, which can be achieved by adjusting the horizontal tune along with the families of sextupoles (harmonic and normal) to tune the amplitude-dependent tune shift (ADTS) to stabilize the beam \cite{bessy:2015}. This approach is effective but empirical tuning of the sextupoles may be required. With this empirical tuning method, TRIBs have been successfully observed at MLS, BESSY-II, and MAX-IV \cite{max4:2019}. Other facilities such as SSRL \cite{spear3:2022} and SLS \cite{paul:2022} have also reported TRIBs observation. So far to our best knowledge, these light sources that have established a second stable orbit (TRIBs) are at relatively low energy ($\le3$~\unit{GeV}). To understand how TRIBs can form at the Cornell Electron Storage Ring (CESR), a relatively high-energy (6 GeV) storage ring with strong radiation damping, we chose a different approach by designing a lattice with the horizontal tune near the 3rd-order integer line and a new sextupole distribution with desired resonant driving terms (RDTs) \cite{bengtsson:1997}, at which TRIBs could form by adjusting the horizontal tune only. Indeed the TRIBs were observed experimentally in our newly designed lattice \cite{suntao:2022}.

The TRIBs lattice design, especially the optimization of sextupoles, is based on a simple equation derived from Hamiltonian perturbation theory \cite{sylee_book, suntao:2022}. This novel practical approach provides an informative guide for studying TRIBs. However, calculation of the TRIBs locations (fixed points in phase space) using the equation from Hamiltonian perturbation theory may not be accurate enough because it relies on knowing accurate coefficients (see discussion in the following section). Thus, a map-based method equipped with FPP-PTC codes utilizing the one-turn resonant map are used to calculate the fixed points in phase space precisely \cite{ef_book:1997, ef_book:2016}. Note FPP-PTC refers to ``Fully Polymorphic Package"-``Polymorphic Tracking code" library which handles Taylor map manipulation and Lie algebraic operations. For simplicity, PTC will be used to refer to the entire package in this paper. PTC tools are general and can handle a system with radiation, which is inevitable in the lepton storage rings with strong radiation damping. In the next section, we will briefly discuss the radiation effect following the introduction of the theory. In addition, a special sextuple knob designed to change corresponding RDTs can adjust the TRIBs location in the horizontal phase space. Experimental results are observed with varying this sextupole knob, which agree reasonably well with the theoretical calculations and simulations, demonstrating a unique method to manipulate the particle's phase space.

In this paper, we will discuss the theoretical calculations of TRIBs near a 3rd-order resonance line with both Hamiltonian perturbation theory and map-based theory \cite{ef_book:2016}, both of which have been well established, and here we will outline the necessary equations for better discussions. Then we will show the experimental results observing TRIBs at various conditions at CESR. The paper is organized as follows. The theory of the nonlinear resonance near a 3rd-order resonance is discussed in Sec.~\ref{theory}. The lattice design and in particular the optimization of sextupoles is discussed in Sec.~\ref{design}. Particle tracking simulations, developed to demonstrate the formation of TRIBs at different tunes with optional clearing kicks, are discussed in Sec.~\ref{sim}. Theoretical calculations of adjustment of TRIB's location in the horizontal phase space are discussed in Sec.~\ref{phase_con}. The experimental results demonstrating the formation and control of TRIBs in phase space are shown in Sec.~\ref{exp}. Finally, the discussion and conclusion are in Sec.~\ref{discussion}.

\section{Theory}\label{theory}
Hamiltonian perturbation theory is normally used to study the particle motion near a resonance as discussed in many textbooks \cite{ef_book:1997, sylee_book, hw_book:2015, ef_book:2016}. Here we start with the normalized Hamiltonian (neglecting radiation for simplicity), the initial theoretical input of any calculation, to show the analytic formula of the fixed points in the action-angle phase space near the $3\nu_{x}$ resonance. When the horizontal tune of an accelerator is near the 3rd-order resonance ($3\nu_{x}=l$) at which the potential formation of islands is expected, the particle's normalized Hamiltonian in the lowest order is 
\begin{align}
H_r &= \delta J_{x} + \nu_y J_{y} + \frac{1}{2}\alpha_{xx}J_{x}^2  + \frac{1}{2}\alpha_{yy}J_{y}^2 + \alpha_{xy} J_{x}J_{y} \nonumber \\ 
    &+ |G|J_{x}^{3/2}cos(3\phi_{x}+\phi_{0})  \textrm{,} \label{eq:h} 
\end{align}
where $J_x$ and $\phi_x$ are the conjugate phase-space coordinates (action and angle), $G$ ($|G|e^{i\phi_0}$) is the complex coefficient of the resonance strength at the 3rd-order resonance 3$\nu_x=l$, and $\delta=\nu_{x}-\frac{l}{3}$, where $\nu_{x}$ is the horizontal tune. $J_y$ and $\nu_{y}$ are the particle's action and tune in the vertical plane. $\alpha_{xx}$, $\alpha_{yy}$ and $\alpha_{xy}$ are the detune coefficients of ADTS. Note (x, y) denote the normalized variables in two planes, which do not necessarily refer to the horizontal and vertical planes in real space (e.g. under coupling).

Indeed, the form of $H_r$ in Eq.~(\ref{eq:h}), which does not depend on $s$ (longitudinal coordinate), is a pure statement about the topology of phase space (three islands appearance), independent of the techniques we use to compute. It is written down at the lowest order without any calculations but the coefficients remain to be computed. For example, the term with $G$ is necessary for the formation of islands. If $G=0$ for an accelerator with superperiodicity, no islands in phase space will be observed until the symmetry is broken such that $G\neq 0$ \cite{bessy:2022}.

The coefficients $\alpha_{xx}$ and $G$ can be derived from purely analytic procedures in terms of accelerator parameters such as sextupole strengths, phase advances and the lattice functions. But this is very difficult even for the simplest model of an accelerator \cite{ef_book:1997}  (see Appendix~\ref{appenAna} for such expressions). However, assuming that Eq.~(\ref{eq:h}) is true, it is then obvious that the one-turn Lie map (in Dragt's notation) at the position of observation $s$ must be given by \cite{ef_book:1997, ef_book:2016}
\begin{equation}
{{\cal M}}_{s}={{\cal A}}_{s}^{-1}\exp\left({:-{2\pi l \over 3}{J}_{x}:}\right)\exp\left({:-2\pi {H}_{r}:}\right){{\cal A}}_{s}\ \label{eq:map}  \textrm{,}
\end{equation}
where ${\cal M}_{s}$ is the one-turn map and ${\cal A}_{s}^{-1}$ is the $s$-dependent canonical transformation that was used to get the Hamiltonian as Eq.~(\ref{eq:h}). The exponential operator is defined as 
\begin{equation}
\exp\left({:f:}\right) \overset{\mathrm{def}}{=} \sum_{k=1} ^{\infty} \frac{:f:^{k}}{k!} \textrm{,}
\end{equation}
where $:f:$ is the Lie operator and $:f:^{0}\overset{\mathrm{def}}{=}$Identity. Besides Ref.~\cite{ef_book:1997, ef_book:2016}, other relevant contribution to the analytical treatment of Hamiltonian using normal form can be found in Ref.~\cite{ef_paper:1989, ef_paper:1990, bazzani:1993, cern:1994}. 

Knowing the Hamiltonian, the fixed points are then determined by the Hamilton's equations with the following conditions
\begin{equation}
\frac{dH_r}{dJ_x} = 0 \textrm{,}\ \ \ \ \frac{dH_r}{d\phi_x} = 0 \textrm{.}
\label{eq:fix}
\end{equation}
Assuming the average $<J_y>$ due to radiation is much less than $J_x$ at the fixed points, and solving Eq.~(\ref{eq:fix}), the action and angle of the fixed points are found. These are then the island centroids:
\begin{align}
J_{x}^{1/2} &= (-1)^{k+1} \frac{3|G|}{4\alpha_{xx}} (1\pm\sqrt{1-\frac{16\alpha_{xx}\delta}{9G^{2}}} )\textrm{,} \label{eq:action} \\
\phi_x &= \frac{k\pi-\phi_0}{3} \textrm{,} \label{eq:angle}
\end{align}
where $k = \pm1,3$ and $k = \pm2,0$ for either three stable fixed points (SFPs) or unstable fixed points (UFPs). The higher order effects in $J_x$ can be handled properly with the full ${\cal A}_{s}$ in the PTC calculation. Equation~(\ref{eq:action}) implies there are four possible solutions for $J_x^{1/2}$ but only two solutions are physical because $J_x^{1/2}>0$. Therefore, one solution of $J_x^{1/2}$ is for SFPs (except zero orbit) and the other for UFPs. If the solution of the action of the SFPs ($J_{\mathrm{x SFP}}$) is either too small ($\sim$0) or too large (exceed the physical aperture), no TRIBs will be observed. If two solutions of $J_x^{1/2}$ require different $k$ ($k_1 = \pm1,3$ and $k_2 = \pm2,0$), three SFPs separated by $\frac{2\pi}{3}$ will be intercalated with three UFPs by $\frac{\pi}{3}$ in the normalized phase space. We name this type TRIBs as the first type. If two solutions of $J_x^{1/2}$ have the same $k$ ($k_{1,2} = \pm1,3$ or $\pm2,0$), three SFPs will have the same angle as three UFPs in the normalized phase space determined by Eq.~(\ref{eq:angle}). This is the second type of TRIBs. These two types of fixed point contours have been recognized early in mathematics \cite{henon:1969} and are recently mentioned in an accelerator paper \cite{max4:2021}. The first type with intercalated SFPs and UFPs is the most common one, observed in CESR and many other facilities. In Sec.~\ref{design}, we will discuss how to design these two types of TRIBs at CESR. Equation~(\ref{eq:action}) also implies that the formation of TRIBs depends on three variables $|G|$, $\alpha_{xx}$, and $\nu_x$. Thus, empirically adjusting the horizontal tune ($\nu_x$) and the ADTS ($\alpha_{xx}$) helps the formation of TRIBs. Moreover, Eq.~(\ref{eq:angle}) indicates the island locations in phase space can be controlled by adjusting the phase $\phi_0$ of $G$, which will be demonstrated in detail in Sec.~\ref{phase_con}.

The analytic formula of the action and angle of the fixed points (Eq.~(\ref{eq:action}) and (\ref{eq:angle})) provides a direct guide to study the TRIBs such as for lattice design and phase space manipulation. However, these action-angle solutions are not exact but approximate (higher order and coupling are neglected) as we will see when comparing the tracking results in Sec.~\ref{sim}. Finding accurate solutions requires knowing accurate parameters $G$ and $\alpha_{xx}$ in advance, which sometimes are difficult to obtain. Therefore, a numerical method is developed using PTC \cite{ef_book:1997,ef_book:2016} to find the exact solutions. It follows from Eq.~(\ref{eq:map}) that a normal form of the one-turn Taylor map provides the coefficients $\alpha_{xx}$, $\delta$ and $G$ in the most general case and that this can be extended to higher order effortlessly. Equivalent to Eq.~(\ref{eq:action}) and (\ref{eq:angle}), the action and angle of the fixed points ($J_{xn}$, $\phi_{xn}$) can be solved from the numerical Hamiltonian. Then the estimated solutions are used as a seed for a Newton search to find the exact solutions of the fixed points \cite{ef_book:2016}. In Sec.~\ref{sim} we will demonstrate this map-based method and compare with tracking results. Also, the code equipped with the Truncated Power Series Algebra (TPSA) package \cite{berz_book:1999}, such as PTC, can use sextupole strengths as parameters and get the parametric dependence of Eq.~(\ref{eq:h}) to any order. In Appendix~\ref{appenNvx}, we show the calculation of fixed points near a $4\nu_x$ or $5\nu_x$ resonance line using this map-based method.

As stated earlier, radiation is neglected in the above discussion. Actually, PTC can handle the system with radiation \cite{ef_book:2016}. We have implemented PTC codes both with and without including radiation. The difference between the results (fixed point positions) including radiation and without (Hamiltonian) are in the 6-7\% range, consistent with tracking results. We agree that under certain circumstances, the calculation including radiation might indeed be necessary. However, it does not seem to be the case in our present application as discussed in the following sections. In any event, there is no big coding differences between the two cases in PTC and the validity of a Hamiltonian calculation can be easily checked and have been. The theory of which PTC handles the radiation is included in Appendix~\ref{appenRad}.

\begin{table}[t]
   \centering
   \caption{CESR Parameters for CHESS-U operation}
   \begin{tabular}{lcr}
   \hline
Beam energy (GeV)             &$E_0$              & $6.0$\\
Circumference (m)             &$L$                &$768.438$\\
Transverse damping time (ms)  &$\tau_{x,y}$             &$12.0$, $14.6$\\
\quad\quad\quad\quad\quad\quad\quad\quad\quad\quad\quad(turns)  &                &4685, 5700\\
Longitudinal damping time (ms)  &$\tau_z$             & $8.2$\\
Horizontal tune               &$\nu_x$              & $16.556$\\
Vertical tune                 &$\nu_y$              & $12.636$\\
Synchrotron tune              &$\nu_s$              & $0.027$\\
Horizontal emittance (nm rad)  &$\epsilon_x$     &$\sim28$\\
Energy spread                  &$\sigma_p$        & $8.2\times10^{-4}$\\     
Revolution frequency (kHz)          &$f_{rev}$         & $390.14 $  \\
   \hline
   \end{tabular}
   \label{table1}
\end{table}

\section{Lattice design}\label{design}

CESR is a 6-GeV accelerator located on the Cornell University campus. Since 2008, CESR has served as a dedicated light source for x-ray users, namely Cornell High Energy Synchrotron Source (CHESS). In 2018, one sextant of the ring was upgraded with double bend acromat to reduce the emittance and accommodate more compact undulators \cite{chessu:2019}. The main accelerator parameters after the upgrade (CHESS-U) are listed in Table \ref{table1}. CESR magnets including 113 quadrupoles, 12 dipole quadrupoles (combined function dipoles), and 76 sextupoles are all individually powered, which provides flexibility for lattice design and complex nonlinear dynamics studies. Unlike most 3rd and 4th generation light sources, there is no global periodicity in the CESR lattice. Details of our CESR lattice can be found in Ref. \cite{chessu:2019}.

\begin{figure}
   \centering
   \includegraphics*[width=240pt]{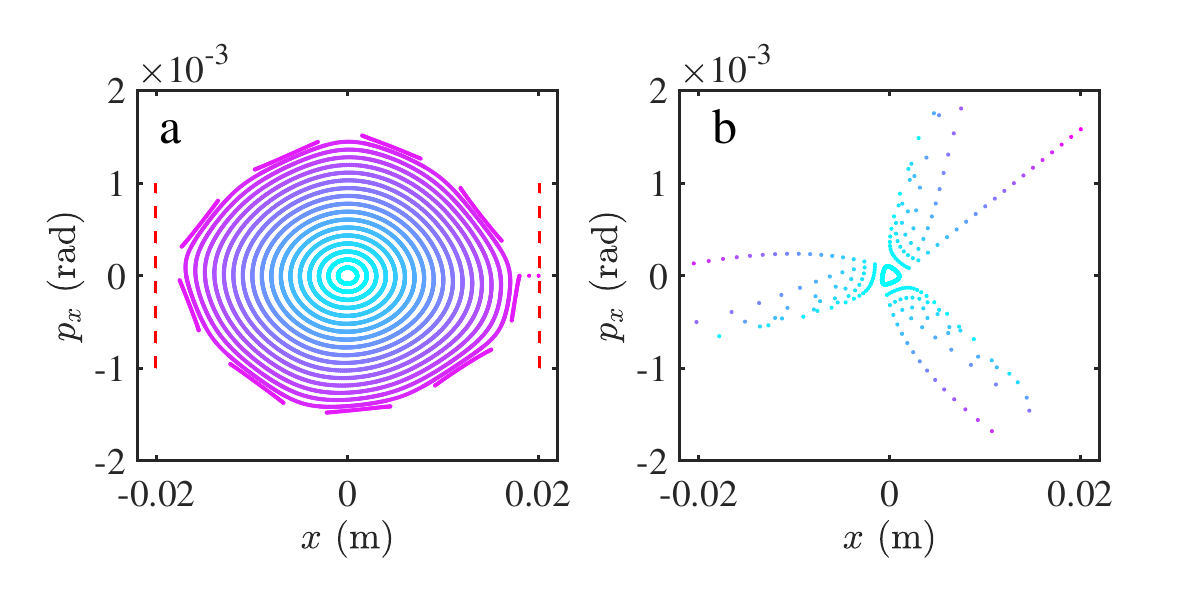}
   \caption{Particles' horizontal phase space at (a) $Q_x=0.556$ and (b) 0.661 in the CHESS-U lattice. The red dashed lines indicate the projected horizontal physical aperture.}
   \label{fig:des_lat}
\end{figure}

As shown in Table \ref{table1}, the nominal horizontal fractional tune ($Q_x$) during CHESS operation is $0.556$, far from the 3rd-order resonance ($\sim$0.667). It is very unlikely that TRIBs would appear by only adjusting the tune to near 0.667. In CESR with standard CHESS-U lattice, we explored adjusting the horizontal tune from 0.556 to near the 3rd-order resonance but the tuning resulted in beam loss without observing TRIBs. This is then confirmed in tracking simulation, where 20 particles with different initial actions are tracked through the CHESS-U lattice at two $Q_x$ (0.556 and 0.661) for 1000 turns while RF, radiation damping and excitation are turned off. As Fig.~\ref{fig:des_lat} (a) shows, at the lattice design tune most of particles (18 out of 20) survive 1000 turns, indicating good horizontal dynamic aperture (DA). While at $Q_x=0.661$ near the 3rd-order resonance, only one particle with the smallest action survives, indicating very poor DA (Fig.~\ref{fig:des_lat} (b)). Note here three UFPs are visible in this phase plot but not SFPs. Thus, changing $Q_x$ to near the 3rd-order resonance alone without adjusting the sextupoles seems impossible for TRIBs to appear in the CHESS-U lattice. A new lattice is needed.

To avoid the difference resonance ($Q_x=Q_y$) for later tune adjustment, we first optimize quadrupoles to create a linear lattice with fractional tunes at (0.643, 0.579) closer to the 3rd-order line while preserving most optics parameters the same as the normal CHESS-U lattice. Then 76 sextupoles are optimized to meet the conditions of forming TRIBs (Eq.~(\ref{eq:action})) as well as maximizing the DA \cite{wb_thesis:2020}. The basic optimization procedure is to vary the strength of sextupoles to minimize a merit function including two chromaticities and all the RDTs with Levenberg-Marquardt least-squares method.

For this sextupole optimization, appropriate target values of $G$ and $\alpha_{xx}$ are needed. Suppose the horizontal tune approaches the 3rd-order line from below and TRIBs appear at $Q_x=0.654$ ($\delta=-0.013$) with $J_{SFP}=1.0\times10^{-5}$~\unit{m\ rad}, which is within the range that our visible-light beam size monitor (vBSM) can measure the TRIBs \cite{suntao:2013}. The range of vBSM means the source region limits that the instrument can image, which is determined by the detector sensor size and the magnification factor of the optics. Since $\delta$ is negative, positive $\alpha_{xx}$ will satisfy the bifurcation condition $\frac{16\alpha_{xx}\delta}{9G^2}\leq1$. Using Eq.~(\ref{eq:action}) with the above requirements, we obtain one set of values $|G|=0.1$~m$^{-\frac{1}{2}}$ and $\alpha_{xx}=1347$~m$^{-1}$ for the sextupole optimization. Then for optimization constraints, the horizontal and vertical chromaticities are constrained to 1, and the target value of $|h_{30000}|$ was set to 0.1~m$^{-\frac{1}{2}}$ and the target of $h_{22000}$ (real number) was set to $-1347$~m$^{-1}$ while other RDTs are minimized. All the RDTs are defined in Ref.~\citep{bengtsson:1997, wang:2012}. As a reminder, we also discuss the RDTs briefly in Appendix~\ref{appenRDT}. $h_{30000}$ is the coefficient of the resonant mode $(h_x^{+})^3(h_x^{-})^0(h_y^{+})^0(h_y^{-})^0\delta_e^0$, where $h_x^{\pm}=\sqrt{2J_x}e^{\pm i\phi_x}$ and $h_y^{\pm}=\sqrt{2J_y}e^{\pm i\phi_y}$ are the resonance basis and $\delta_e$ is the energy offset \cite{bengtsson:1997}. Since this resonant mode and its complex conjugate containing $J_x^{3/2}$ describe the resonance at $3\nu_x$, its coefficient $h_{30000}$ is used to approximate $G$ in the optimization. Similarly, $-h_{22000}$ is used to approximate $\alpha_{xx}$. More specifically, $G=2h_{30000}$ and $\alpha_{xx}=-2h_{22000}$ (see Appendix~\ref{appenRDT}). Actually, $-2h_{22000}$ is part of $\alpha_{xx}$ and becomes the leading term when ${\cal A}_{s}$ is mostly linear. (see Sec.~\ref{discussion} for details). Also in the early stage of optimization, the factor 2 was omitted for initial trials as indicated above. After the sextupole optimization, final values of $|h_{30000}|$ and $h_{22000}$ are $0.06$~m$^{-\frac{1}{2}}$ and $-1310$~m$^{-1}$, respectively. The negative $\delta$ with positive $\alpha_{xx}$ leads to $1-\frac{16\alpha_{xx}\delta}{9G^2}>1$, which results in different $k$ for SFPs and UFPs. Thus, the TRIBs in this lattice fall in the first type as discussed in Sec.~\ref{theory}.

For the second type of TRIBs, Eq.~(\ref{eq:action}) implies $0<\frac{16\alpha_{xx}\delta}{9G^2}<1$ so that positive $\alpha_{xx}$ requires $\delta>0$. In the above TRIBs lattice, adjusting $Q_x$ to above $\frac{2}{3}$ will make $\delta>0$. However, the bifurcation condition $\frac{16\alpha_{xx}\delta}{9G^2}\leq1$ is not satisfied any more so that no TRIBs will be observed when $Q_x>\frac{2}{3}$ in this lattice. Suppose $\alpha_{xx}=1000$~m$^{-1}$ and $\delta=0.005$, the bifurcation constraint leads to $|G|>3$~m$^{-\frac{1}{2}}$ and $J_x<6.25\times10^{-5}$~\unit{m\ rad}. With these constraints applied in the sextupole optimization, a second TRIBs lattice is obtained to demonstrate the second type of TRIBs. After optimization, final values of $|h_{30000}|$ and $h_{22000}$ at $Q_x$=0.6691 are $3.35$~m$^{-\frac{1}{2}}$ and $-988$~m$^{-1}$, respectively. As discussed above, the conditions of forming the second type of TRIBs are stricter than the first type. This is probably the reason that the first type of TRIBs is commonly observed in most facilities. In this paper, we mainly focus on the lattice with the first type of TRIBs unless explicitly indicated.

\begin{figure}[t]
   \centering
   \includegraphics*[width=230pt]{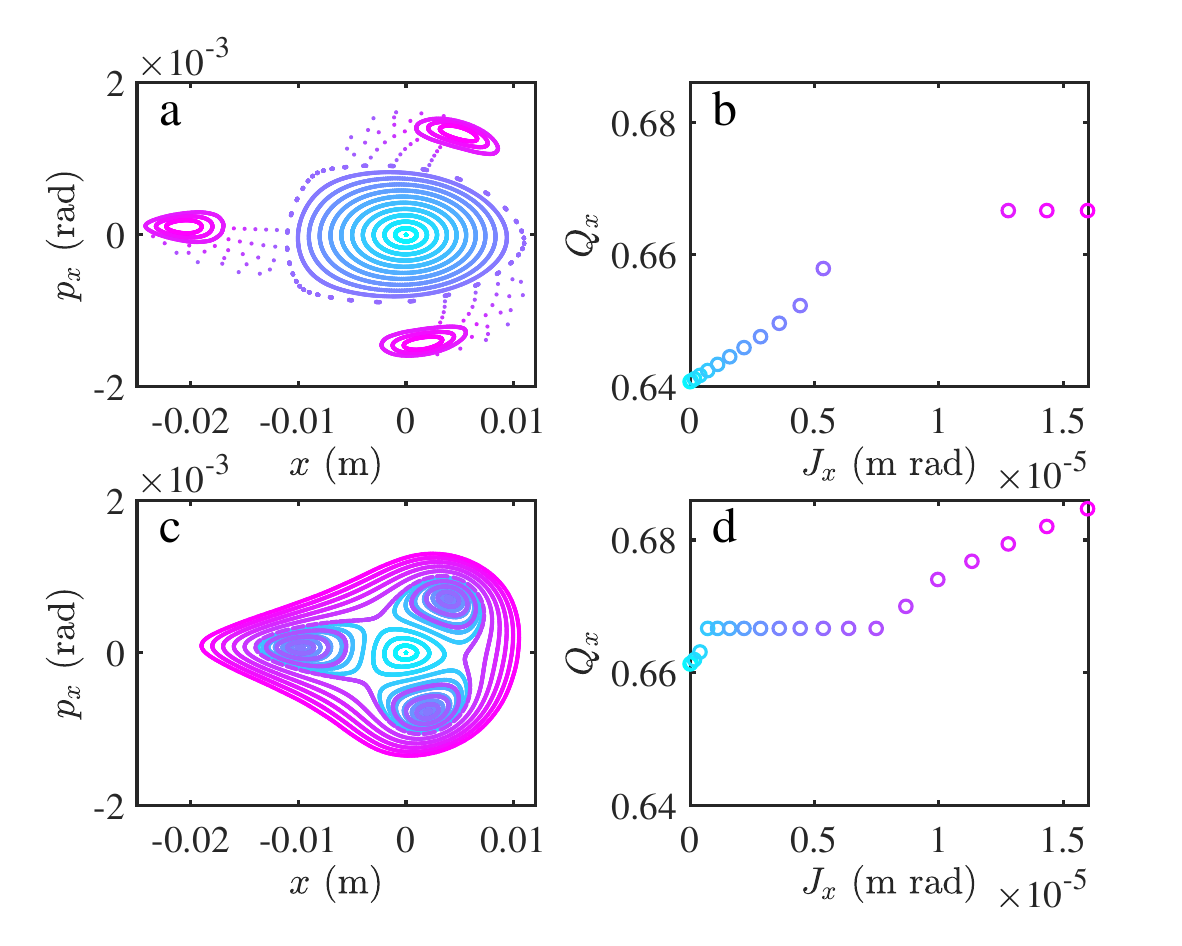}
   \caption{Particles' horizontal phase space at (a) $Q_x=0.643$ and (c) $0.661$. The particle's $Q_x$ as a function of its action in (b) and (d) are calculated from the TBT data plotted in (a) and (c), respectively.}
   \label{fig:chessu}
\end{figure}

\section{Simulation}\label{sim}

\begin{figure}
   \centering
   \includegraphics*[width=240pt]{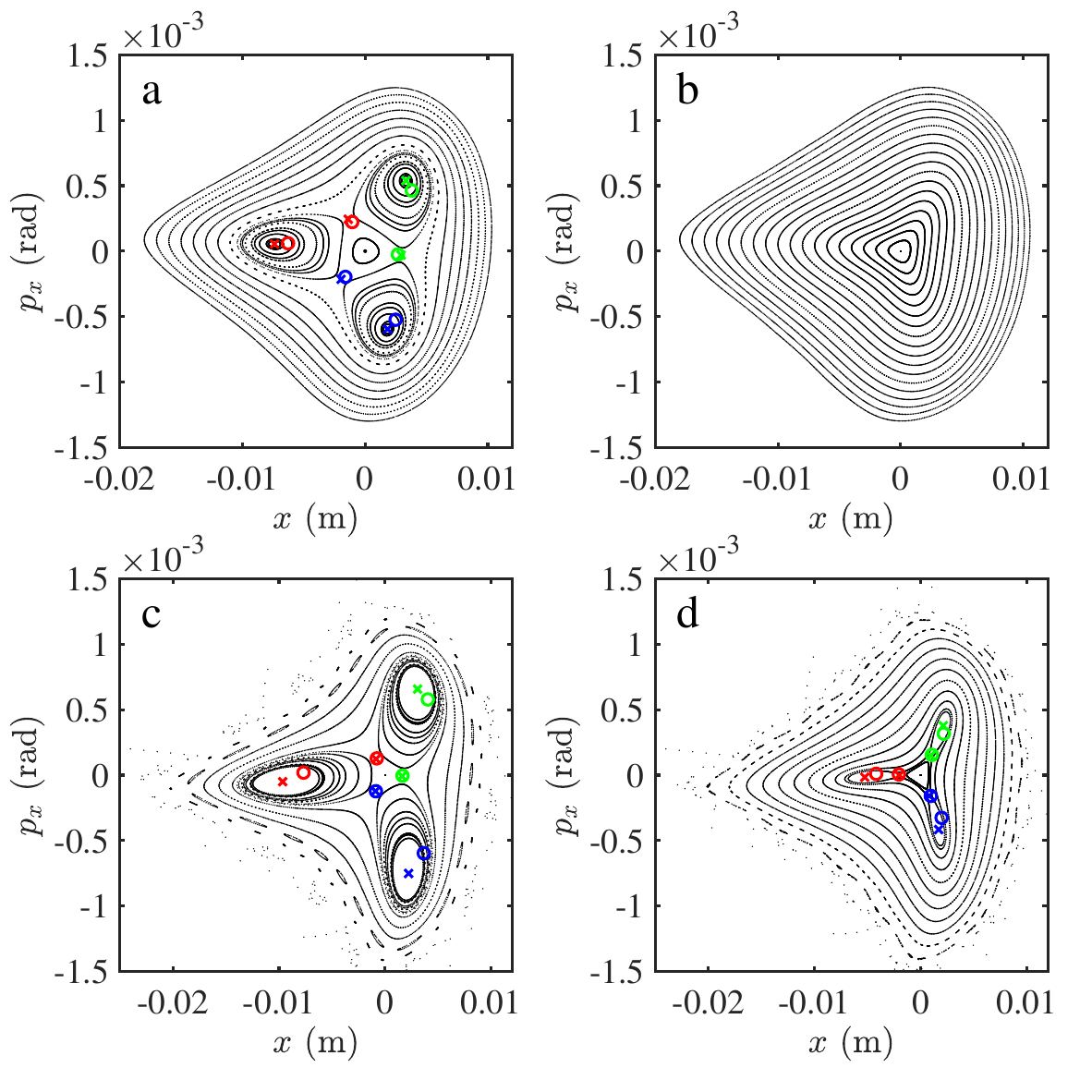}
   \caption{Particles' horizontal phase space from the first type TRIBs lattice at (a) $Q_x=0.664$ and (b) $0.668$ and from the second type TRIBs lattice at (c) $Q_x=0.665$ and (d) $0.668$. The black dots are the tracking results. The circles and crosses are the estimated and exact fixed points calculated by PTC codes, respectively.}
   \label{fig:ptc}
\end{figure}

Once the linear lattice is set and the sextupoles are optimized as discussed above, simulations are implemented to check the lattice properties. All the simulation programs discussed in this paper are based on the BMAD code library \cite{bmad:2006}. First, 20 particles with different initial actions are tracked through the TRIBs lattice at different $Q_x$ (0.643 to 0.666) for 1000 turns while RF, radiation damping and excitation are turned off. Since tracking starts at the first element of the lattice, the recorded coordinates ($x$, $p_x$) in the phase plots are at the first element as well just for convenience. Figure~\ref{fig:chessu} (a) and (c) plot the horizontal coordinates of all particles at $Q_x=0.643$ and $0.661$, respectively. Three stable islands are clearly visible in phase space at both tunes. From FFT analysis of the turn-by-turn (TBT) $x$ coordinates, the particles' tunes are calculated and shown as a function of their actions in Fig.~\ref{fig:chessu} (b) and (d). Both plots show that $Q_x$ increases as the particle's action increases, indicating a positive ADTS coefficient ($\alpha_{xx}>0$). From Fig.~\ref{fig:chessu} (b) at the lattice design tune, the ADTS coefficient ($\alpha_{xx}$) when $J_x<0.5\times10^{-5}$~\unit{m\ rad} can be estimated as 2656~m$^{-1}$, which is fairly close to the design value of $-2h_{22000}$ (2620~m$^{-1}$). When the particles are in the stable islands, their tunes are exactly $\frac{2}{3}$. Several particles are lost while tracking through the lattice at $Q_x=0.643$ so that their tunes are absent in Fig.~\ref{fig:chessu} (b).

\begin{figure}
   \centering
   \includegraphics*[width=260pt]{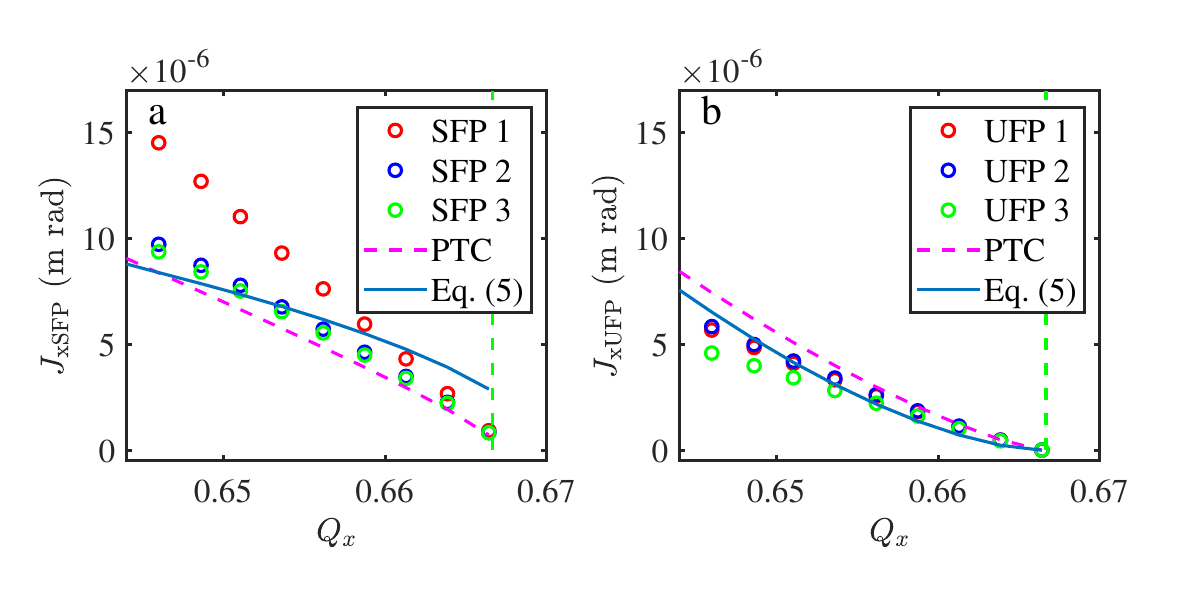}
   \caption{The particle action at the (a) SFP and (b) UFP as a function of lattice tune. The circles and the magenta dashed lines are the exact and estimated actions calculated by PTC codes, respectively. The blue lines are the estimated actions calculated from Eq.~(\ref{eq:action}). The green dashed line indicates the 3rd-order resonance line.}
   \label{fig:fpt}
\end{figure}

As discussed in Sec.~\ref{theory}, PTC codes have been developed to find the action and angle of the fixed points accurately. Figure~\ref{fig:ptc} (a) shows the estimated and exact fixed points of the TRIBs lattice (first type) calculated by PTC codes at $Q_x=0.664$ as well as the tracking results in phase space. Although the estimated fixed points by PTC are already very close to the tracking results, the exact fixed points found by PTC agree even better with the tracking results. For this first type TRIBs lattice, no TRIBs are observed when $Q_x$ is above the 3rd-order resonance line (Fig.~\ref{fig:ptc} (b)). However, in the second type TRIBs lattice, TRIBs formed at tunes below $\frac{2}{3}$ of the first type (Fig.~\ref{fig:ptc} (c)) appear followed by the second type (Fig.~\ref{fig:ptc} (d)) as $Q_x$ increases above $\frac{2}{3}$.

\begin{figure}
   \centering
   \includegraphics*[width=240pt]{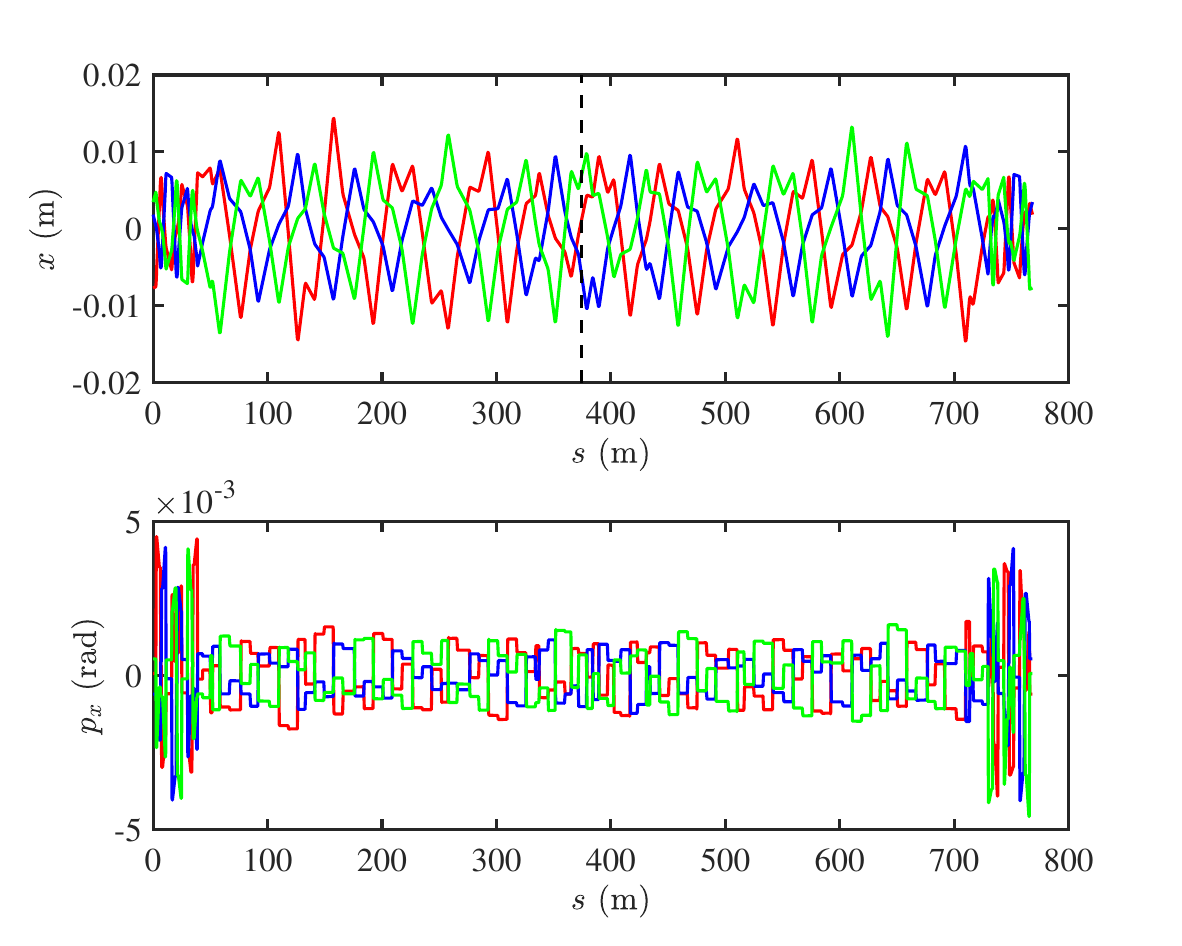}
   \caption{The 3-turn horizontal closed orbit of the particle in the TRIBs at $Q_x=0.6639$. The red, blue, and green lines are the orbits at turn 1, 2, and 3, respectively. The dashed line indicates the source location of vBSM where beam profile images are taken.}
   \label{fig:co}
\end{figure}

The action of the fixed points can be calculated to leading order using the equation $J_x=(\gamma_xx^2+2\alpha_xxp_x+\beta_xp_x^2)/2$, where $\gamma_x$, $\alpha_x$, and $\beta_x$ are the Twiss parameters at the first element. Actions of all the PTC-calculated fixed points (SFPs and UFPs) at different $Q_x$ are then plotted in Fig.~\ref{fig:fpt}. Both $J_{\mathrm{x SFP}}$ and $J_{\mathrm{x UFP}}$ decrease as the tune approaches $\frac{2}{3}$, indicating the island separation in the horizontal plane reduces. The predicted actions of the fixed points by both PTC and Eq.~(\ref{eq:action}) are plotted in Fig.~\ref{fig:fpt} as well. Overall trend of actions predicted by both methods agree reasonably well with tracking results. However, Eq.~(\ref{eq:action}) predicts $J_{\mathrm{x SFP}}$ poorly when $Q_x$ is close to $\frac{2}{3}$. On the other side, PTC codes show better estimates for both SFPs and UFPs when $Q_x$ is close to $\frac{2}{3}$.

Knowing the coordinates of SFPs at the first element found by PTC, the 3-turn closed orbit can be calculated by tracking through all elements as shown in Fig.~\ref{fig:co}. We see the horizontal separation between the 3-turn closed orbit (TRIBs) and the zero closed orbit (core beam) can be as large as 12$\unit{mm}$ at certain locations.

\begin{figure}[t]
   \centering
   \includegraphics*[width=270pt]{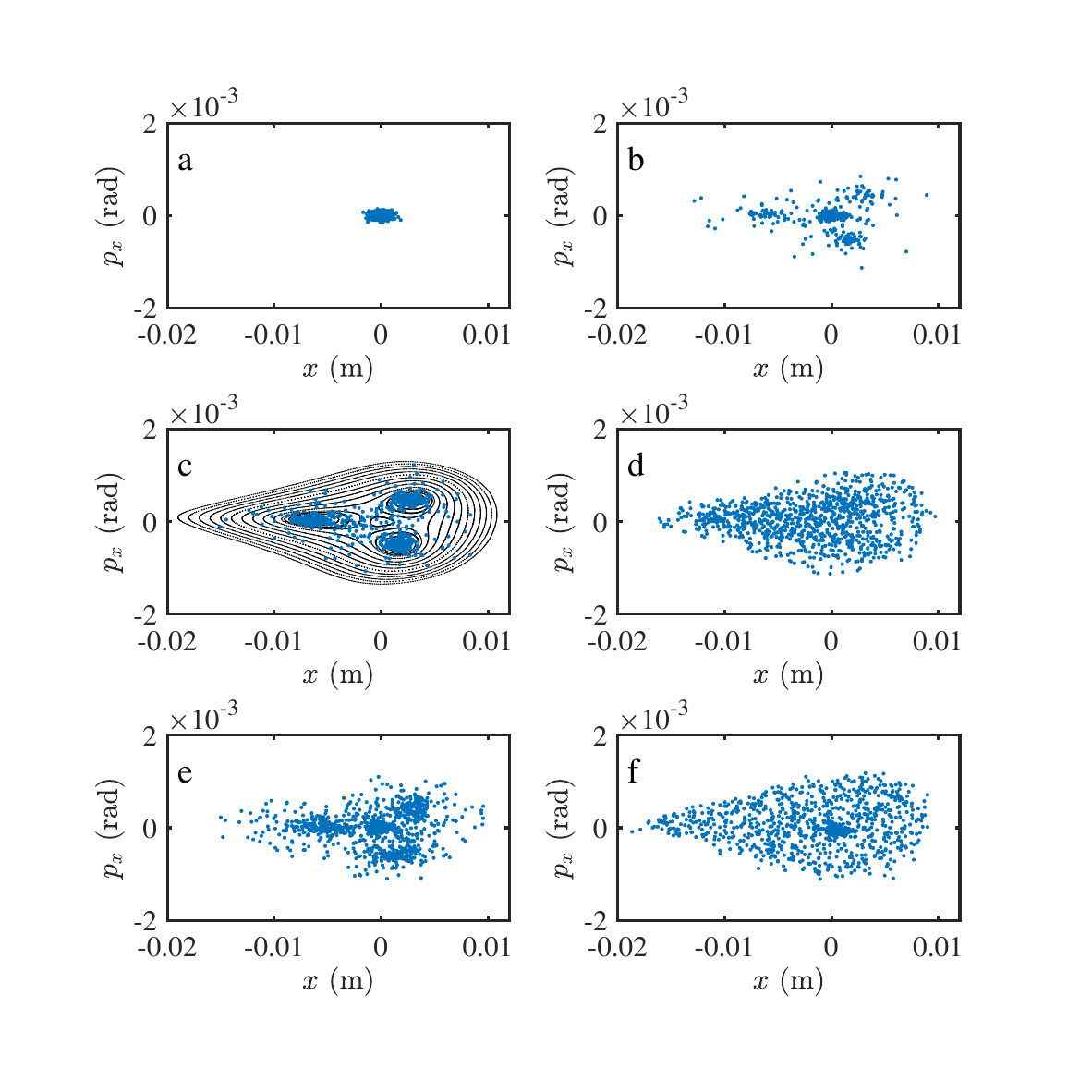}
   \caption{The 1000-particles distribution in the horizontal phase space at $Q_x=0.6649$ (259.4~kHz) at (a) turn 1 and (b) turn 40000, when applying a sinusoidal kick at 259.4~kHz with the kick amplitude of (c) 0.5~$\mu$rad and (d) 5~$\mu$rad at turn 40000, and at 261.35~kHz with the kick amplitude of (e) 0.5~$\mu$rad and (f) 5~$\mu$rad at turn 40000. }
   \label{fig:track}
\end{figure}

The DA of the TRIBs lattice at different tunes has been checked with both tracking and frequency map analysis, which indicate slightly reduced DA compared to the CHESS-U lattice \cite{suntao:2022}. In order to be sure the beam can survive at different tunes, more realistic tracking simulation including radiation damping and excitation is implemented to check the TRIBs formation in the TRIBs lattice. Starting with an initial distribution with design emittance $\epsilon_x=30$~\unit{m\ rad} (Fig.~\ref{fig:track} (a)), 1000 particles are tracked through the lattice at different horizontal tunes. Since the horizontal radiation damping time in CESR is 12~\unit{ms} ($\sim$4700 turns), 4$\times10^{4}$ turns is used for tracking to cover $\sim$8 damping times. When $Q_x<0.6613$ (258~kHz), the particles remain in the similar distribution through 4$\times10^{4}$ turns as seen in Fig.~\ref{fig:track} (a). When $Q_x>0.6639$ (259~kHz), the particles diffuse from the core to the island buckets. Figure~\ref{fig:track} (b) shows the particles occupy three islands as well as the core after tracking 4$\times10^{4}$ turns at $Q_x=0.6649$ (259.4~kHz).

Starting with the distribution in Fig.~\ref{fig:track} (b) and applying a transverse sinusoidal kick at 259.4~kHz same as the core tune, most particles are driven from the core to the islands (Fig.~\ref{fig:track} (c)) after tracking 4$\times10^{4}$ turns. However when the sinusoidal kick amplitude is greater than 3 $\mu$rad, the particles are not clustered in the islands but distributed in the ($x, p_x$) phase space evenly (Fig.~\ref{fig:track} (d)). 

When starting tracking with the initial distribution as in Fig.~\ref{fig:track} (c) with the sinusoidal kick at 261.35~kHz near the island tune (0.667), the particles converge back to the core bucket but are not fully cleared from the islands buckets after 4$\times10^{4}$ turns (Fig.~\ref{fig:track} (e)). When increasing the kick amplitude to 5~$\mu$rad, the particles are cleared from island buckets but most of them are evenly distributed in phase space (Fig.~\ref{fig:track} (f)). This tracking simulation with clearing kicks indicates that there may exist an optimal kick amplitude which drives the particles between the islands and core bucket without particle loss.

In Fig.~\ref{fig:track} (c), the tracking results without including radiation damping and excitation are plotted as well. From this figure, we can see the damped particles are indeed confined to the stable islands, of which locations are consistent with the SFPs predicted by the simulations without including radiation damping and excitation. It is understandable that the radiation will damp the particles to the SFPs (core or islands) but may not change the positions of SFPs much as confirmed by the PTC calculation including radiation damping. Therefore, we only show the tracking results without including radiation damping and excitation in later discussions.

\begin{table}
   \centering
   \caption{RDTs and phase angle changes from four different sextupole distributions. $\phi_0=\tan^{-1}(\frac{h_{30000i}}{h_{30000r}})$, $\Delta\phi_x=-\frac{\Delta\phi_0}{3}$. $\psi_{x1}$ is the angle of the first SFP (red) in the normalized phase plot Fig.~\ref{fig:phi0}.}
   \begin{tabular}{lcccc}
   \hline
Parameters      &Design      &1       &2      &3      \\
                &a     &b       &c      &d     \\
\hline
$h_{30000r}$ (m$^{-\frac{1}{2}}$)      &-1.3427      &-0.9922    &-0.1532      &1.0544 \\
$h_{30000i}$ (m$^{-\frac{1}{2}}$)       &-0.0939    &-0.8911    &-1.2277       &-0.0443  \\
$|h_{30000}|$ (m$^{-\frac{1}{2}}$)       &1.3460   &1.3336    &1.2372       &1.0553  \\
$h_{22000}$ (m$^{-1}$)       &-1529.9      &-1525.2     &-1526.5       &-1520.4  \\
$\phi_0$ (degree)       &-176.00     &-138.07     &-97.12       &-2.41  \\
$\Delta\phi_x$ (degree)  & 0      &-12.64     &-26.26       &-57.86  \\
$\psi_{x1}$ (degree)  & 178.28      &170.77     &161.71       &129.76  \\
$\Delta\psi_{x1}$ (degree)  & 0      &-7.51     &-16.57       &-48.52  \\
   \hline
   \end{tabular}
   \label{table2}
\end{table}

\section{Phase space control}\label{phase_con}
As Eq.~(\ref{eq:angle}) indicates, three SFPs (or UFPs) are separated by $\frac{2\pi}{3}$ in the normalized phase space and the angle $\phi_0$ determines the locations of the fixed points. Therefore, changing $\phi_0$ by $\Delta\phi_0$ while keeping other parameters ($Q_x$, $|G|$, and $\alpha_{xx}$) same will rotate all the fixed points by $\Delta\phi_x$($=-\frac{\Delta\phi_0}{3}$) in the normalized phase space. Two approaches are explored to demonstrate the control of the TRIBs locations in phase space.

\begin{figure}
   \centering
   \includegraphics*[width=240pt]{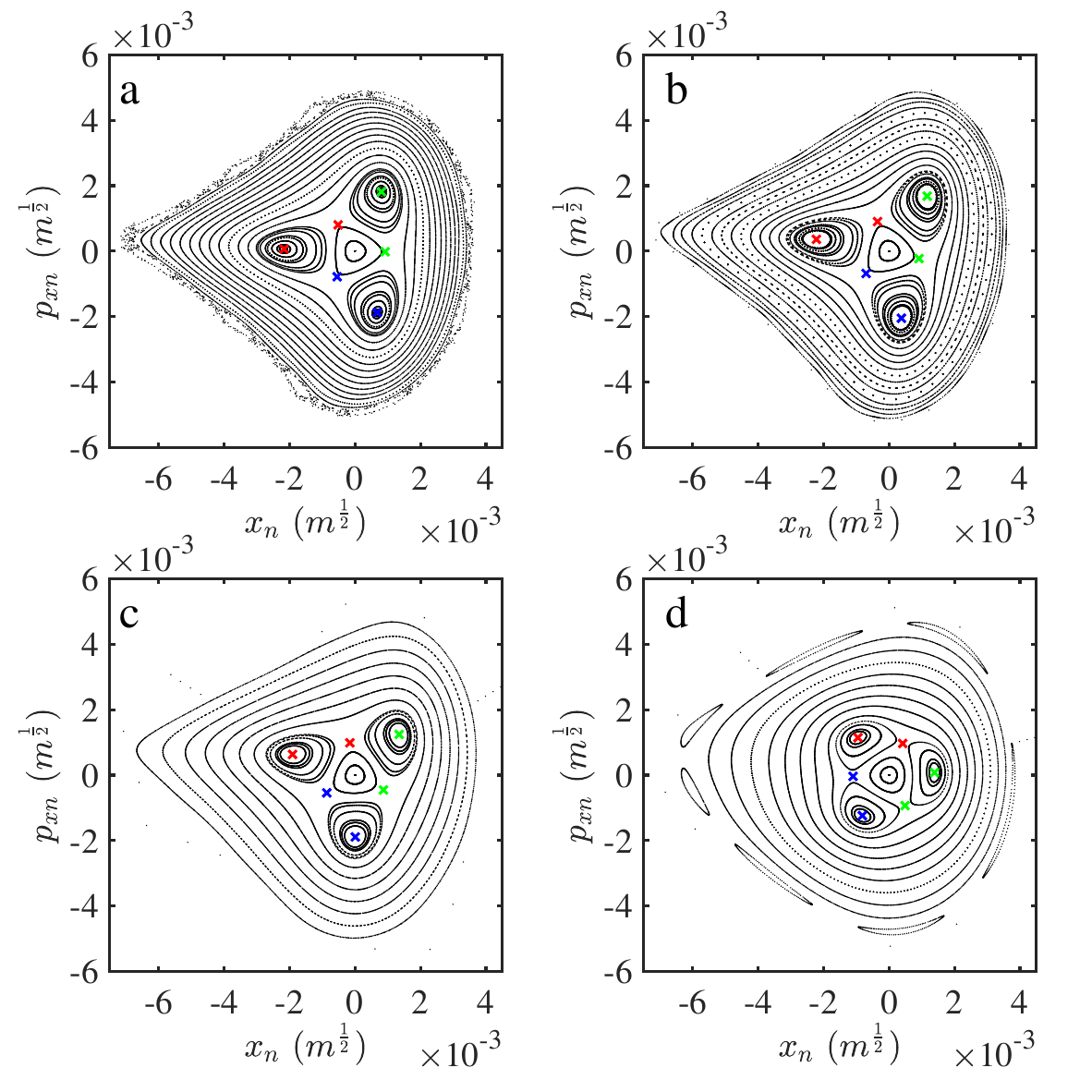}
   \caption{Normalized horizontal phase space from the lattices with (a) the design sextupoles and (b$-$d) three other sextupole distributions at $Q_x=0.6639$. The colored crosses indicate SFPs and UFPs calculated by PTC codes.}
   \label{fig:phi0}
\end{figure}

\subsection{Rotate TRIBs}
Because we use $2h_{30000}$ as the substitute for $G$, the angle $\phi_0$ of $G$ is found by $\phi_0=\tan^{-1}(\frac{h_{30000i}}{h_{30000r}})$, where $h_{30000r}$ and $h_{30000i}$ are the real and imaginary parts of $h_{30000}$. The first approach is to optimize all the sextupoles with targets set for $h_{30000r}$ and $h_{30000i}$ so as to change $\phi_0$ while keeping the change of other RDTs minimum. Three sextupole distributions are then optimized to demonstrate three different phase angle $\phi_0$. Table~\ref{table2} summarizes the related RDTs and calculated phase angles from the design (a) and three other sextupole distributions (b-d). To view and evaluate the phase angle, we plot the normalized phase space of the tracking results from these four sextupole distributions in Fig.~\ref{fig:phi0}. The linearly normalized coordinates ($x_n$, $p_{xn}$) are obtained by
\begin{equation}
\begin{pmatrix}
x_n \\
p_{xn} \\
\end{pmatrix} =
\begin{pmatrix}
\frac{1}{\sqrt{\beta_x}} & 0 \\
\frac{\alpha_x}{\sqrt{\beta_x}}  & \sqrt{\beta_x} \\
\end{pmatrix}
\begin{pmatrix}
x \\
p_{x} \\
\end{pmatrix}
 \textrm{.    }
\label{eq:norm}
\end{equation}
The angles of all the fixed points in the normalized phase space are calculated using  $\psi_{x}=\tan^{-1}(\frac{p_{xn}}{x_{n}})$. As an example, we calculate the angle ($\psi_{x1}$) of the first SFP (red cross in Fig.~\ref{fig:phi0}) as well as the expected angle $\phi_0$ for four sextupole distributions and list them in Table~\ref{table2}. As Fig.~\ref{fig:phi0} shows, the TRIBs rotate clockwise gradually as $\psi_{x1}$ decreases ($\phi_{0}$ increases) from (a) the design lattice to (d) the third sextupole distribution. Shown in Table~\ref{table2}, the expected angle change $\Delta\phi_x$ ($=-\frac{\Delta\phi_0}{3}$) is consistent with the actual angle change $\Delta\psi_{x1}$, which demonstrates the control of the TRIBs location in phase space by varying $\phi_0$. Practically, based on the design and three other sextupole distributions, knobs can be created to vary all the sextupoles so as to adjust the TRIBs locations in phase space from one state to another.

\begin{figure}
   \centering
   \includegraphics*[width=240pt]{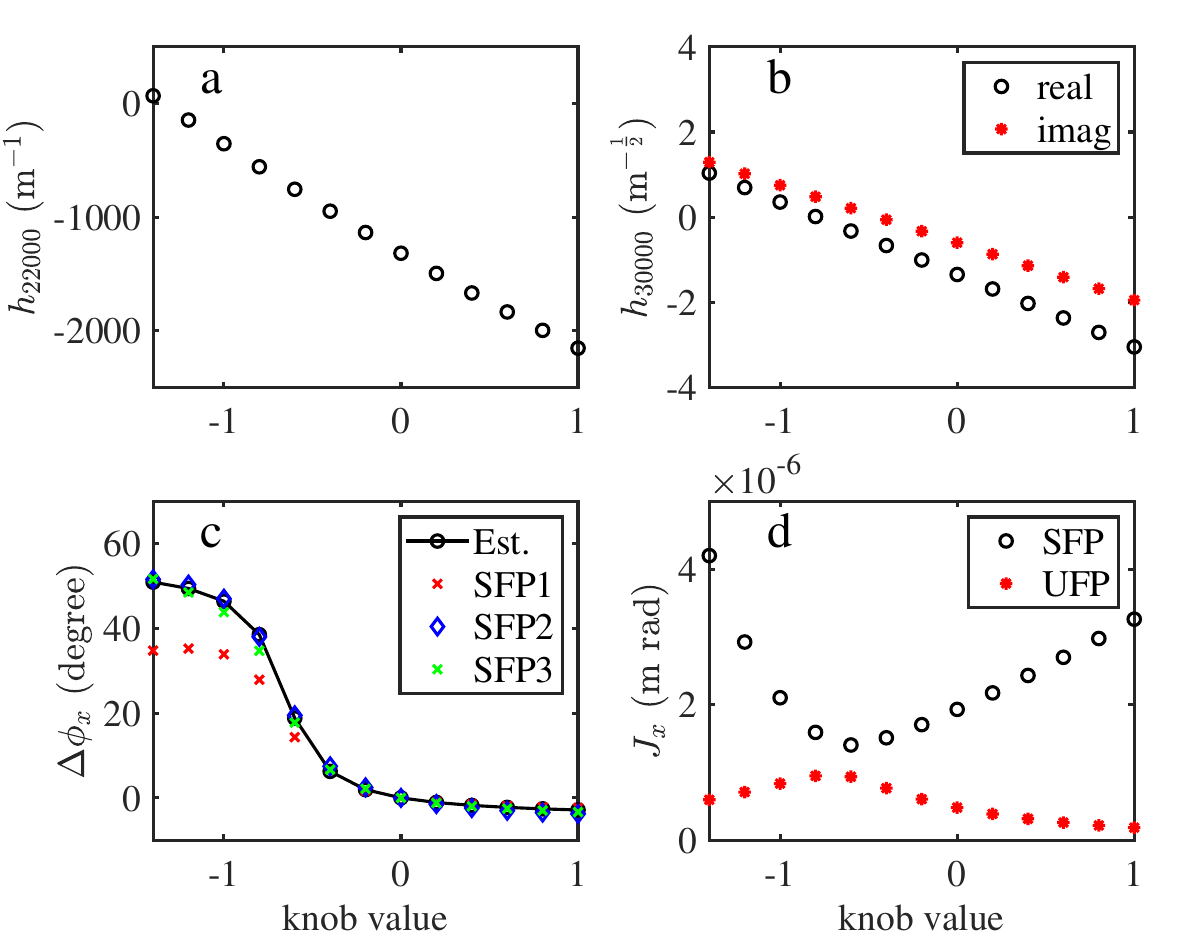}
   \caption{RDT terms ($h_{22000}$, $h_{30000r}$, $h_{30000i}$), estimated and exact phase angle change, and estimated $J_x$ of the fixed points change as function of the sextupole knob value while the lattice tune is kept at $Q_x=0.6639$.}
   \label{fig:knob_para}
\end{figure}

\subsection{Knob control}
\begin{figure}
   \centering
   \includegraphics*[width=240pt]{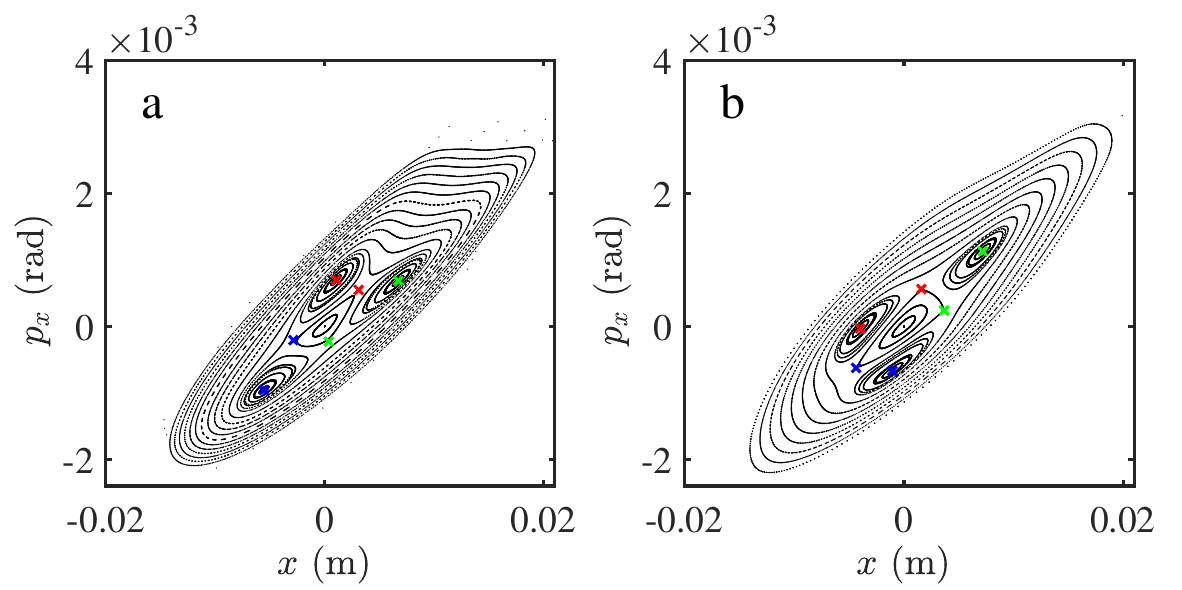}
   \caption{Particles' horizontal phase space at the vBSM source location from two lattices with the sextupole knob value set at (a) 0 and (b) $-1$ at $Q_x=0.6639$. The colored crosses indicate the SFPs and UFPs calculated from tracking.}
   \label{fig:knob_phase}
\end{figure}

In preparation for the experimental test of the TRIBs lattice, we create a knob which will vary $h_{22000}$ at a large scale while keeping $|h_{30000}|$ at minimum change. Five harmonic sextupoles are chosen and grouped to form this sextupole knob. These sextupoles are far away from the locations having large horizontal and vertical orbit offsets in order to avoid large impact on the orbit when changing them. Also by design the change of chromaticity is negligible while adjusting the knob. As shown in Fig.~\ref{fig:knob_para} (a) and (b), when the knob value changes from $-1.2$ to 1.0, $h_{22000}$ shows a large linear change from $\sim0$ to $-2100$ but less change for both $h_{30000r}$ and $h_{30000i}$ (1 to -2). Interestingly, although the absolute value changes of both $h_{30000r}$ and $h_{30000i}$ are small, they yield large angle change of the fixed points in phase space. As shown in Fig.~\ref{fig:knob_para} (c), the estimated angle change ($\Delta\phi_x=-\frac{\Delta\phi_0}{3}$) gradually increases to $\sim50^o$ when the knob value changes to $-1.2$, indicating the fixed points rotate counterclockwise in phase space (Fig.~$\ref{fig:knob_phase}$). The exact phase angle change of three SFPs at different knob values are calculated from normalized phase plots (similar as Fig.~$\ref{fig:phi0}$) and plotted in Fig.~$\ref{fig:knob_para}$ (c). The exact angle changes of SFP2 and SFP3 agree very well with the estimate but there exists discrepancy for SFP1. This angle discrepancy is not understood yet, possibly due to larger nonlinear effect with slightly larger $J_x$ at SFP1. Besides the angle, the estimated actions of SFPs and UFPs change along with the knob value as shown in Fig.~$\ref{fig:knob_para}$ (d).

Figure~$\ref{fig:knob_phase}$ (a) and (b) show the particle's ($x$, $p_x$) at the vBSM source location with the sextupole knob value set at 0 and $-1$, respectively. As we can see, dialing the knob from 0 to $-1$ will rotate the TRIBs in phase space and the x-ray spots of the TRIBs in $xy$ real space will change accordingly. The horizontal coordinates of three SFPs at the vBSM source point are $x_1=1.03\unit{mm}$ (red), $x_2=-5.58\unit{mm}$ (blue), and $x_3=6.71\unit{mm}$ (green) with the knob value set at zero, and $x_1=-3.99\unit{mm}$, $x_2=-1.06\unit{mm}$, and $x_3=7.18\unit{mm}$ with the knob value set at $-1$. These expected horizontal coordinates from PTC calculation at different sextupole knob values will be compared with the experimental results in the following section.

\begin{figure}
   \centering
   \includegraphics*[width=220pt]{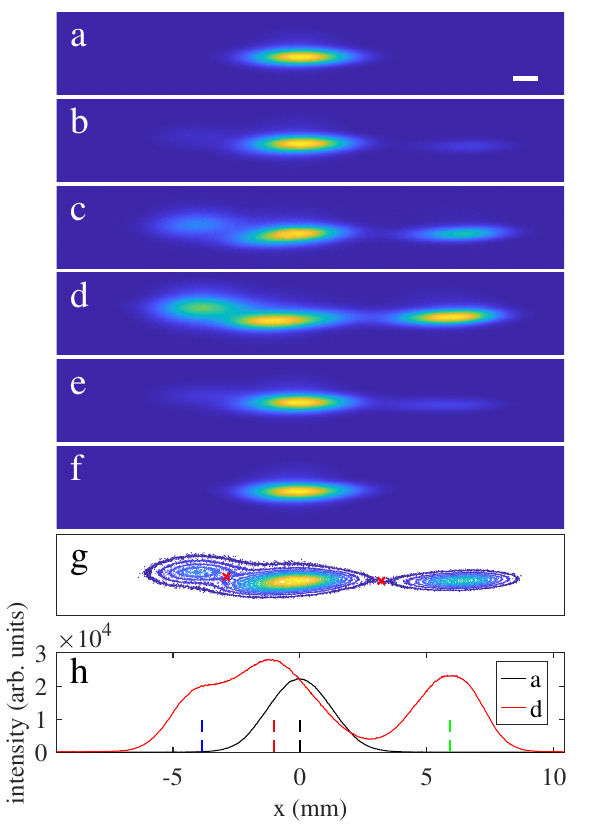}
   \caption{Beam images at $Q_x$= (a) 0.6613 (258.0~kHz), (b) 0.6631 (258.7~kHz), (c) 0.6648 (259.35~kHz), and (f) 0.6700 (261.4~kHz). (d) and (e) are recorded at the same tune 0.6648 (259.35~kHz) but with applying a sinusoidal kick at 259.35~kHz and 261.35~kHz, respectively. (g) is the contour plot of image (c). Two red crosses indicate two UFPs. (h) shows the horizontal profile of images (a) and (d). The dashed lines indicate the centers of the core and  three islands. The white line indicates a 1-mm scale.}
   \label{fig:firstexp}
\end{figure}

\section{Experiment}\label{exp}

After the confirmation of TRIBs in simulation, we loaded the TRIBs lattice into CESR for experimental studies. The optics correction including orbit, betatron phase, coupling, and dispersion were first made at the design horizontal tune ($Q_x=0.643$) \cite{betatron:2000}. Then we adjusted $Q_x$ from 0.6537 to 0.667 while viewing and recording the beam profile from synchrotron radiation using vBSM \cite{suntao:2013}. The beam profile images from a 1-mA single positron bunch observed at different tunes are shown in Fig.~\ref{fig:firstexp}. When $Q_x<0.663$ (258.7~kHz), the beam stays in the core (Fig.~\ref{fig:firstexp} (a)). When $Q_x$ is near 0.663 (258.7~kHz), the TRIBs start to form (Fig.~\ref{fig:firstexp} (b)) while most particles are still in the core bucket. At $Q_x=0.6648$ (259.35~kHz), more particles diffuse to the island buckets (Fig.~\ref{fig:firstexp} (c)). A bunch-by-bunch feedback \cite{dimtel} was then used to apply a transverse sinusoidal kick ($\sim$0.1 $\mu$rad) with the same frequency as the core tune 259.35~kHz to the beam. This clearing kick drives the particles from the core to the island buckets (Fig.~\ref{fig:firstexp} (d)). When the sinusoidal kick's frequency was set to 261.35~kHz, this clearing kick drove the particles from the islands to the core (Fig.~\ref{fig:firstexp} (e)). But some particles remain in the island buckets. At $Q_x=0.670$ (261.4~kHz) above the 3rd-order resonance line, all the particles diffuse back to the core island (Fig.~\ref{fig:firstexp} (f)). The behavior of the clearing kick is also consistent with the results reported in Ref. \cite{bessy:2015}. 

From these images, the horizontal positions of the core and three stable islands can be extracted. For example, the horizontal profiles of the images in Fig.~\ref{fig:firstexp} (a) and (d) are plotted in Fig.~\ref{fig:firstexp} (h) (offset by the core position). The horizontal coordinates of the beam at the core and three islands are then found by fitting the peaks with a Gaussian function, and the distances between three islands and the core are $x_1=-1.03\unit{mm}$, $x_2=-3.84\unit{mm}$, and $x_3=5.91\unit{mm}$, respectively, which are in good agreement with the SFPs calculated by PTC and tracking: $x_1=-1.50\unit{mm}$, $x_2=-3.42\unit{mm}$, and $x_3=5.95\unit{mm}$. To estimate the UFPs from the experimental results, the image in Fig.~\ref{fig:firstexp} (c) is shown as contours in Fig.~\ref{fig:firstexp} (g) for better viewing the UFPs. Clearly, two UFPs are present in this plot. Their horizontal positions are estimated as $-2.9\unit{mm}$ and $3.2\unit{mm}$, different from the calculated UFPs, $-3.27\unit{mm}$ and $2.34\unit{mm}$. This implies that the features of the ($x$, $p_x$) phase plot cannot be reflected completely accurately in the vBSM images in the $xy$ plane.

\begin{figure}
   \centering
   \includegraphics*[width=220pt]{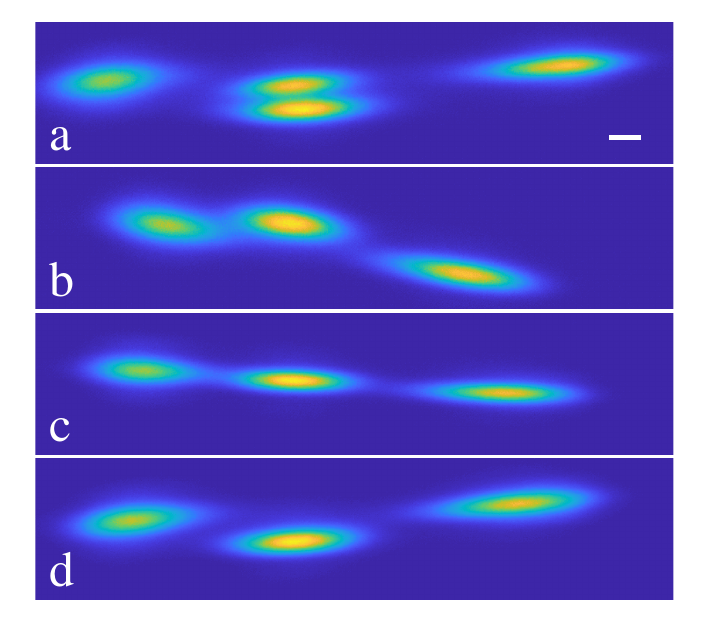}
   \caption{Beam profiles (a) at $Q_x=0.6646$ (259.3\unit{kHz}) with a skew quadrupole at +10000\unit{cu}, and at $Q_x=0.6664$ (260.0\unit{kHz}) with the skew quad strength set at (b) $-$10000, (c) 0, and (d) +10000\unit{cu}, respectively. The white line indicates a 1-mm scale.}
   \label{fig:coupling}
\end{figure}

\begin{figure}
\centering
\includegraphics*[width=220pt]{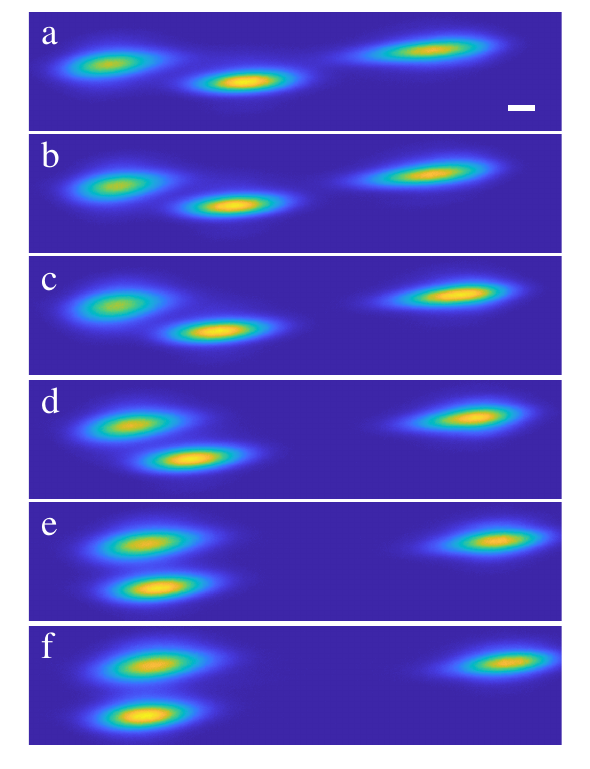}
\caption{Beam profiles with the sextupole knob set at (a) 0 (b) $-$200, (c) $-$400, (d) $-$600, (e) $-$800 and (f) $-$900\unit{cu}, respectively. The white line indicates a 1-mm scale.}
\label{fig:knob_exp}
\end{figure}

When the $xy$ coupling is corrected and small ($<2\%$), three TRIBs observed in the $xy$ plane will be along a line near $y=0$ as observed in Fig.~\ref{fig:firstexp} (d). If the particles occupy both the core and the TRIBs, the image of the center core will overlap with one of the TRIBs near the center in the $xy$ plane so that they are hardly distinguishable (Fig.~\ref{fig:firstexp} (c)). To understand how the coupling affects the separation between the core beam and the TRIBs in the $xy$ plane, an experiment was conducted by varying the strength of a skew quadrupole ($sk\_q14w$) so as to change the $xy$ coupling while viewing the TRIBs with a 2-mA single positron bunch in CESR.  As shown in Fig.~\ref{fig:coupling} (a) recorded at $Q_x=0.6646$ (259.3\unit{kHz}) with the skew quadrupole at an increased strength $\Delta K_1L=5.74\times10^{-3}$ m$^{-1}$ (10000~\unit{cu}) with the measured $xy$ coupling of $\sim8\%$, the core and the TRIBs are coexisting but well separated in the $xy$ plane. Here ``cu" refers to ``computer unit", the digital unit in the real machine. To further demonstrate the coupling effect, we varied the strength of $sk\_q14w$ from $-5.74\times10^{-3}$ m$^{-1}$ ($-$10000\unit{cu}) to $5.74\times10^{-3}$ m$^{-1}$ (10000\unit{cu}) in a step of 2000\unit{cu} to change the $xy$ coupling gradually at $Q_x=0.6664$ (260.0\unit{kHz}). Only three islands, no core beams, were observed at this tune. The images during this process were recorded in a video (See Supplemental Material movie 1). Figure~\ref{fig:coupling} (b), (c) and (d) show the recorded images at three different strengths of $sk\_q14w$:  $-5.74\times10^{-3}$, 0 and $5.74\times10^{-3}$ m$^{-1}$, respectively, which clearly demonstrates the particle's $y$ coordinate depends on its ($x$, $p_x$) coordinates because of the $xy$ coupling. When the polarity of $sk\_q14w$ changes, the coupling term including $p_x$ may change its sign as well. Thus, the center island beam in Fig.~\ref{fig:coupling} (b) and (d) is above and below the core beam, respectively. The increased coupling leads to the increase of vertical beam size, which can be clearly seen in Fig.~\ref{fig:coupling} (b) and (d) comparing to Fig.~\ref{fig:coupling} (c). Note here Fig.~\ref{fig:coupling} (c) is slightly different from Fig.~\ref{fig:firstexp} (d) because of slightly different sextupole distribution in the two experiments. As a side effect, the TRIBs images provide an interesting visual diagnostic tool to display the coupling and for potential coupling measurement.

\begin{figure}
\centering
\includegraphics*[width=200pt]{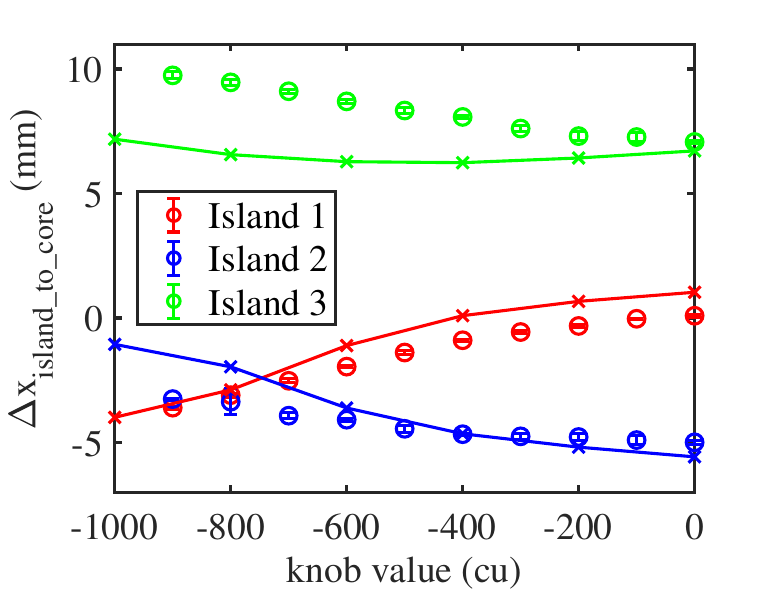}
\caption{Stable islands locations extracted from experimental images (circles) and simulation (crosses) as a function of the sextupole knob value.}
\label{fig:knob_exp_ana}
\end{figure}

For demonstration of the control of the TRIBs location in phase space, the sextupole knob discussed in Sec.~\ref{phase_con} was loaded into CESR control knobs. Dialing the sextupole knob by $1000$\unit{cu} in the real machine corresponds to changing the knob value by 1 in the simulation (Fig.~\ref{fig:knob_para}). The knob was adjusted gradually in a step of $-100$\unit{cu} at $Q_x=0.6664$ (260.0\unit{kHz}). The images during the adjustment were recorded in a movie (See Supplemental Material movie 2). The beam profile recorded at the various knob values of 0, $-200$, $-400$, $-600$, $-800$, and $-900$\unit{cu} are shown in Fig.~\ref{fig:knob_exp} (a) to (f), respectively. As the knob value decreases, as shown from Fig.~\ref{fig:knob_exp} (a) to (f), the center TRIB moves towards negative $x$ and away from the center, eventually passing the second TRIB at $-900$\unit{cu} (Fig.~\ref{fig:knob_exp} (f)). This is consistent with the simulation results in Fig.~\ref{fig:knob_phase}. One can clearly see TRIBs rotate around the center (the core beam) while adjusting the sextupole knob. From the images in Fig.~\ref{fig:knob_exp}, the horizontal positions of three islands relative to the core are extracted and plotted in Fig.~\ref{fig:knob_exp_ana} as well as the calculated SFPs by PTC. We can see the stable island locations from experimental results are roughly consistent with the simulation results (Fig.~\ref{fig:knob_phase}). The larger discrepancy between the experiment and simulation results at higher knob values may come from the fact that during experiment the sextupole knob decreased the tune slightly, which separates one island (green) more from the core and complicates the TRIBs locations in the $xy$ plane.

\begin{figure}
\centering
\includegraphics*[width=220pt]{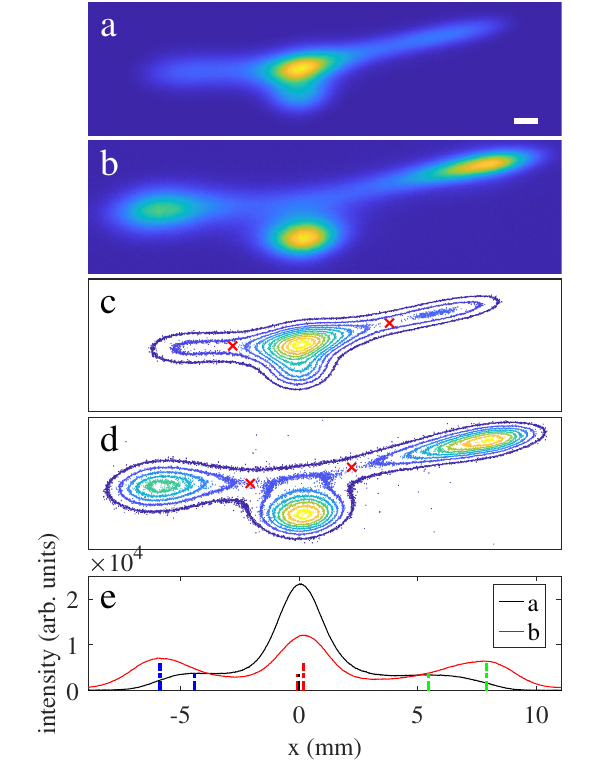}
\caption{Beam images observed at (a) $Q_x=0.6674$ (260.4\unit{kHz}) and (b) 0.6654 (259.6\unit{kHz}) in the second TRIBs lattice. (c) and (d) are the contour plots of images (a) and (b), respectively. The red crosses indicate the UFPs. (e) shows the horizontal profiles of images (a) and (b). The dashed lines indicate the center locations of core and islands. The white line indicates a 1-mm scale.}
\label{fig:type2_exp}
\end{figure}

To check how differently the second type of TRIBs behave, we loaded the second type TRIB lattice in CESR. After optics corrections were made at higher tune ($Q_x=0.672$) above the 3rd-order resonance, the beam profile images of a single 1-mA positron bunch were recorded continuously while lowering the horizontal tune. The TRIBs start to appear around $Q_x=0.668$ (260.6\unit{kHz}). As $Q_x$ decreases to 0.6674 (260.4\unit{kHz}) still above the resonance line, the TRIBs are well separated from the core but connected to the core as shown in Fig.~\ref{fig:type2_exp} (a). When $Q_x$ continues to decrease below $\frac{2}{3}$, the core starts to disappear, more particles are diffused into islands, and three TRIBs are further separated horizontally. As shown in Fig.~\ref{fig:type2_exp} (b), at $Q_x=0.6654$ (259.6\unit{kHz}), the particles occupy all three islands but long tails exist in between. TRIBs are observed at tunes both below and above the 3rd-order resonance, which is consistent with the simulation shown in Fig.~\ref{fig:ptc} (c) and (d) and suggests the first and second type of TRIBs are observed below and above the resonance line respectively in this TRIBs lattice. Furthermore, Fig.~\ref{fig:type2_exp} (c), the contour plot of the image in Fig.~\ref{fig:type2_exp} (a), shows the similar contours as in Fig.~\ref{fig:ptc} (d), suggesting the second type of TRIBs is observed. Two UFPs extracted from the contour plots in Fig.~\ref{fig:type2_exp} (c) and (d) listed in Table~\ref{table3} are very different from the calculated UPFs from simulation. However, the horizontal positions of stable islands extracted from the horizontal profiles (Fig.~\ref{fig:type2_exp} (e)) of the images in Fig.~\ref{fig:type2_exp} (a) and (b) are in good agreement with the results calculated by PTC as listed in Table~\ref{table3}.

\begin{table}
\centering
\caption{The horizontal positions of SFPs and UFPs extracted from experimental results (Fig.~\ref{fig:type2_exp}) and simulation.}
\begin{tabular}{lccccccc}
\hline
Qx   &  &SFP1      &SFP2       &SFP3      &UFP1 &UFP2 &UFP3      \\
     &  &(mm)      &(mm)       &(mm)      &(mm) &(mm) &(mm)      \\
\hline
0.6674 &Exp  &-0.08      &-4.44    &5.47     &      &-2.80 &3.81     \\
       &Sim  &-0.11      &-4.11    &4.99     &-0.15 &-1.62 &1.89     \\
0.6654 &Exp   &0.17      &-5.87    &7.91     &      &-2.07 &2.21     \\
      &Sim   &0.32      &-6.57    &8.32     &0.17 &-1.18  &1.17     \\
\hline
\end{tabular}
\label{table3}
\end{table}

\section{Discussion and conclusion}\label{discussion}
We have briefly discussed the Hamiltonian perturbation theory near a $3\nu_x$ resonance line and the resulting equations of the action and angle of the fixed points in phase space. These equations (Eq.~(\ref{eq:action}) and (\ref{eq:angle})) provide necessary conditions for TRIBs to form in a storage ring. Based on these equations and using RDTs ($|h_{30000}|$ and $h_{22000}$), we successfully designed several lattices to form TRIBs and experimentally observed TRIBs in these lattices at CESR. We have used $-2h_{22000}$ as the substitute of $\alpha_{xx}$ in the sextupole optimization. This procedure was applied before Nishikawa's involvement. While writing this paper, Nishikawa pointed out $h_{22000}$ defined in Ref.~\cite{wang:2012} is part of $\alpha_{xx}$. The exact $\alpha_{xx}$ under the assumptions of Bengtsson-Wang~\cite{bengtsson:1997, wang:2012} can be found in Chapter 5 and 9 in Ref.~\cite{ef_book:1997} and in Appendix~\ref{appenAna}. When $Q_x$ is near the 3rd-order resonance, only if ${\cal A}_{s}$ is mostly linear, could $\alpha_{xx}$ be replaced by $-2h_{22000}$ (the leading term). During the optimization of sextupoles in our TRIBs lattice, the sextupole distribution is optimized by minimizing all RDTs except $h_{30000r}$, $h_{30000i}$ and $h_{22000}$, with the result that they are about 10 or 20 times less than $h_{22000}$, so that ${\cal A}_{s}$ is nearly linear. This is probably why the measured ADTS coefficient ($\alpha_{xx}$) from tracking is consistent with the $-2h_{22000}$ as shown in Sec.~\ref{sim}. When $|h_{22000}|$ is small or comparable to other second order RDTs, $-h_{22000}$ can not represent $\alpha_{xx}$/2 and we should use more accurate method such as map-based method with PTC codes for the lattice design and optimization. As demonstrated in Sec.~\ref{sim}, the map-based method with PTC codes produces accurate estimate and exact calculation of the fix points in phase space. We are in the process of incorporating the PTC codes into the lattice design and optimization program for better TRIBs prediction.

We have also demonstrated both in simulation and experimentally that the TRIBs location in the ($x, p_x$) phase space can be precisely controlled by adjusting a knob grouped with five sextupoles. The experiment results agree reasonably well with the simulation. In addition to the tune adjustment, a combined use of this sextupole knob with a coupling control knob including skew quadrupoles, the TRIBs location especially the x-ray spots from these TRIBs can be easily manipulated in the $xy$ space, which provides huge flexibility for utilizing the x-rays from the island beam in possible timing experiments. 

Both the first and second types of TRIBs have been observed in the simulations and experimentally with newly designed lattices in CESR. As discussed in Sec.~\ref{theory}, much stricter conditions are required to form the second type than the first type, which results in a narrow tune window for the second type experimentally. In addition, possibly due to the locations of UFPs, separation between the core and three islands for the second type are not as clean as the first type, which makes its real machine usage less suitable in light source facilities.

The experiments at CESR were conducted in different runs. Some differences were observed from run to run mainly because the machine conditions were slightly different. In some runs, at the beginning, optics correction (orbit, phase, and coupling) are necessary. After correction, the machine conditions (mainly orbit and coupling) are different from previous runs. The orbit difference shifts the core and island locations in the observed images a little. The horizontal chromaticity sometimes needs adjustment. The different coupling changes the islands separation vertically as we see in Fig.~\ref{fig:coupling}. But in general, the TRIBs behave similarly as expected in different runs. For example, stable islands show up at nearly the same horizontal tune and the phase space control of islands using the sextupole knob is very reproducible.

Our practical method to design and manipulate the islands in phase space has been successfully applied to a 1D third-order resonance. Could this method be extended to study higher order resonances? For example, similar to Eq.~(\ref{eq:action}) and (\ref{eq:angle}), the action and angle of the island centroids near a $n$th-order resonance ($n\geq4$) can be obtained (Appendix~\ref{appenNvx}). Utilizing a similar approach to that described above (or with map-based method), controlling the islands in phase space near the $n$th-order resonance will be straightforward to implement. On the other hand, as the resonance order goes higher, other resonances may interfere with the current resonance so that the one-resonance theory fails and the method could be inapplicable. More importantly, higher order resonances are intrinsically weaker than lower order resonances such that the effects of radiation damping and excitation may not be negligible and then the Hamiltonian method would break down. We found that for the third order resonance studied in this paper, particles close to the resonance islands are lost in times ranging from 30 to 100 turns. These times are nearly two orders of magnitude shorter than the radiation damping time of 4700 turns and explains why the Hamiltonian approach works for this resonance. This must also be true for the method to be applicable to higher order resonances. There are methods to extract a particle loss time, averaged over phase space, in the absence of radiation damping. If this time is larger than or even comparable to the damping time, then the Hamiltonian formalism will break down. Therefore care must be taken when applying the Hamiltonian method for higher order resonances. 

In conclusion, we have demonstrated a new approach to design and observe TRIBs in a high-energy storage ring, which could provide useful guidance for systematic TRIBs design and study. 

\begin{acknowledgments}
The authors thank David Rubin for valuable discussion and comments on this paper, Robert Meller for useful discussion and assistance with the tune tracker, David Sagan for his assistance with BMAD, Joel Brock and Ernest Fontes for supporting this project, P. Goslawski (HZB, Germany) for providing useful information, and CESR operation group for their support during machine study shifts. This research was supported by NSF award PHYS-1757811 and DMR-1829070.
\end{acknowledgments}

\appendix

\section{Islands near a $n\nu_x$ resonance}\label{appenNvx}

Similar to the 3rd-order resonance (3$\nu_x$=$l$),  the particle's normalized Hamiltonian near the $n$th-order resonance ($n\nu_x$=$l$) with expected islands formation in the lowest order is 
\begin{align}
H_r &= \delta J_{x} + \nu_y J_{y} + \frac{1}{2}\alpha_{xx}J_{x}^2  + \frac{1}{2}\alpha_{yy}J_{y}^2 + \alpha_{xy} J_{x}J_{y} \nonumber \\ 
    &+ |G_{n0}|J_{x}^{\frac{n}{2}}cos(n\phi_{x}+\phi_{0})  \textrm{,} \label{eq:h4nu} 
\end{align}
where $G_{n0}$ ($|G_{n0}|e^{i\phi_0}$) is the complex coefficient of the resonance strength at the $n$th-order resonance $n\nu_x$=$l$, $\delta$=$\nu_{x}-\frac{l}{n}$, and $n$ is an integer.

Similarly, the one-turn Lie map at the position of observation $s$ is given by
\begin{equation}
{{\cal M}}_{s}={{\cal A}}_{s}^{-1}\exp\left({:-{2\pi l \over n}{J}_{x}:}\right)\exp\left({:-2\pi {H}_{r}:}\right){{\cal A}}_{s}\ \label{eq:map4nu}  \textrm{,}
\end{equation}
where ${\cal M}_{s}$ is the one-turn map and ${\cal A}_{s}^{-1}$ is the $s$-dependent canonical transformation that was used to get the Hamiltonian $H_r$ as Eq.~(\ref{eq:h4nu}).

With this Hamiltonian, the conditions of fixed points (Eq.~(\ref{eq:fix})) lead to
\begin{align}
\delta + \alpha_{xx}J_x + (-1)^k \frac{n}{2} |G_{n0}|J_{x}^{\frac{n}{2}-1}=0  \textrm{,} \label{eq:jnnu} \\
n \phi_x + \phi_0 = k\pi \textrm{,} \label{eq:phinnu} 
\end{align}
where $k$ is an integer. The action and angle of the fixed points can then be found by solving the above two equations. By examining Eq.~(\ref{eq:jnnu}), we see the equation is a polynomials of degree $n-2$ for $\sqrt{J_{x}}$  (odd $n\ge5$) or a polynomials of degree $\frac{n}{2}-1$ for $J_x$ (even $n\ge4$). The analytic solutions of linear and quadratic equations are well known. The root of a general cubic equation can also be found analytically according to Cardano's formula. For higher order, the roots of polynomials can be found numerically.

\begin{figure}
   \centering
   \includegraphics*[width=220pt]{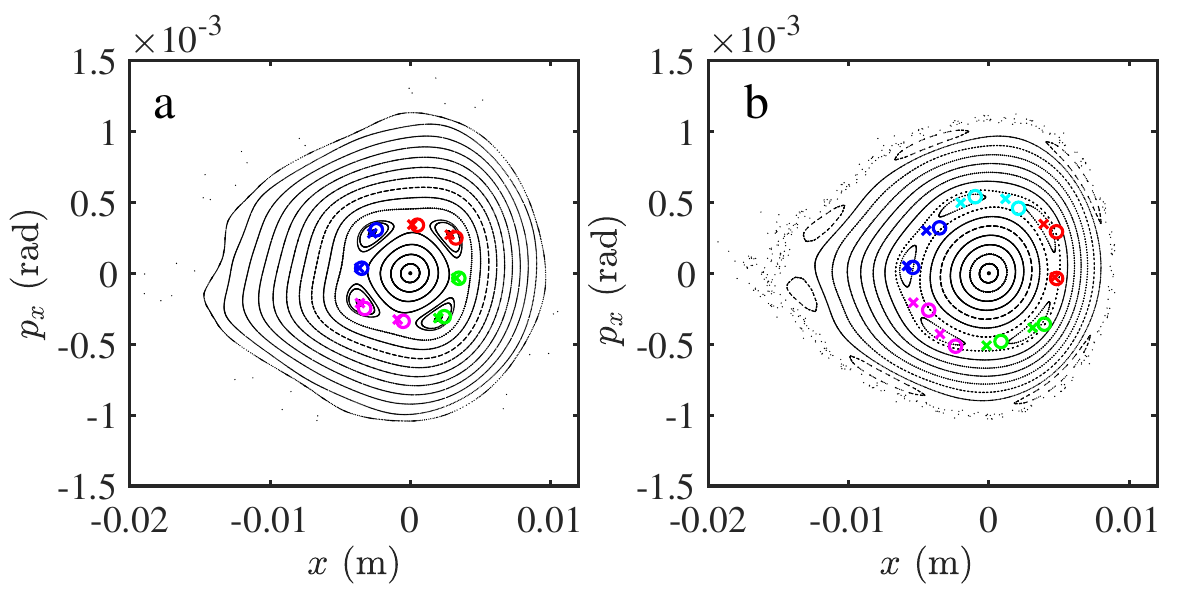}
   \caption{Particles' horizontal phase space from the first type TRIBs lattice at (a) $Q_x=0.7472$ and (b) $Q_x=0.792$. The black dots are the tracking results. The circles and crosses are the estimated and exact fix points calculated by PTC codes, respectively.}
   \label{fig:4nu}
\end{figure}

As a demonstration, we employed the map-based method using PTC codes to calculate the fixed points in the horizontal phase space near a $4\nu_x$ ($Q_x=0.7472$) and a $5\nu_x$ resonance ($Q_x=0.792$) in the first TRIBs lattice. As shown in Fig.~\ref{fig:4nu}, the calculated positions of all the fixed points agree very well with the tracking results.

\section{Radiation effect}\label{appenRad} 
In this paper, we perform a semi-analytical method via TPSA tools in the PTC package. These tools are general and can handle a system with radiation. This is explained in Ref.~\cite{ef_book:2016} where all the calculations are done with general vector fields and thus include radiation if present in the underlying tracking code. It is implemented in BMAD via PTC. The reader is invited to look at section 5.5 of Ref.~\cite{ef_book:2016} where the theory used in this paper is applied to compute a limit cycle in the presence of extremely nonlinear damping.

In the general case, the operator ${H}_{r}$ in PTC is actually of the form: 
\begin{eqnarray} 
:-2\pi {H}_{r}:\ \ \longrightarrow  F_r\cdot \nabla  \textrm{.}
\end{eqnarray}
As pointed out by Bazzani et al.~\cite{bazzani:1993} and also by Forest \cite{ef_book:1997, ef_book:2016}, the one-resonance normal form belongs to a certain gauge group. In the language of Forest used in PTC, the normal form $H_r$ belong to a Lie algebra containing driving terms of the resonance and tune shifts. Unlike the non-resonant normal form, this gauge group is not commutative. This implies that $H_r$ is not unique. However, as pointed out in Ref. \cite{bazzani:1993}, the ``interpolated Hamiltonian'' in the original variable is.  This means that $K_r$ defined from Eq.~(\ref{eq:inter}) is unique:

\begin{align} 
{\cal M}_{s} &= {\cal A}_{s}^{-1}\exp\left({:-{2\pi l \over 3}{J}_{x}:}\right)\exp(:-2\pi {H}_{r}:){\cal A}_{s} \nonumber \\
 &=\exp\left({:-{2\pi l \over 3}{I}_{r}:}\right)\exp\left({:-2\pi {K}_{r}:}\right) \textrm{,} \label{eq:inter}  
\end{align}
where $K_{r}={\cal A}_{s}^{-1}{H}_{r}$ and $I_{r}={\cal A}_{s}^{-1}J_x$.
This fundamental result applies to the non-symplectic case with damping. Thus we have

\begin{align} 
{{\cal M}}_{s} &= {{\cal A}}_{s}^{-1}\exp\left({:-{2\pi l \over 3}{J}_{x}:}\right)\exp\left({{F}_{r}\cdot \nabla }\right){{\cal A}}_{s}\nonumber \\
&=\exp\left({:-{2\pi l \over 3}{I}_{r}:}\right)\exp\left({:{\Phi }_{r}\cdot \nabla :}\right) \textrm{,} \label{eq:interf}
\end{align}
where ${\Phi }_{r}\cdot \nabla ={{\cal A}}_{s}^{-1}{F}_{r}\cdot \nabla {{\cal A}}_{s}$. 

The vector field ${\Phi }_{r}$, also unique in the resonant case, can be easily computed from $F_r$ using the canonical transformation in Taylor form of the Lie map ${\cal A}_s$.  The actual normal form $F_r$ also belongs to a non-commutative algebra which is closed under Lie bracket. Just for the record, it looks like the following expression and we can see that damping terms appear:
\begin{eqnarray}
F_r\cdot \nabla \ &=& T{h}_{x}^{+}{\partial \hphantom{{h}_{x}^{+}} \over \partial {h}_{x}^{+}}+\overline{T}{h}_{x}^{-}{\partial \hphantom{{h}_{x}^{+}} \over \partial {h}_{x}^{-}} \nonumber \\  
&&+S{{h}_{x}^{+}}^{2}{h}_{x}^{-}{\partial \hphantom{{h}_{x}^{+}} \over \partial {h}_{x}^{+}}+\overline{S}{h}_{x}^{+}{{h}_{x}^{-}}^{2}{\partial \hphantom{{h}_{x}^{+}} \over \partial {h}_{x}^{-}}\nonumber \\
 &&+\Gamma {{h}_{x}^{-}}^{2}{\partial \hphantom{{h}_{x}^{+}} \over \partial {h}_{x}^{+}}+\overline{\Gamma }{{h}_{x}^{+}}^{2}{\partial \hphantom{{h}_{x}^{+}} \over \partial {h}_{x}^{-}}+\cdots \textrm{,} \label{eq:fgrad}
\end{eqnarray}
where $T = {d}_{x}+i{\mu }_{x}$, $S={d}_{xx}+{i \over 2}{\alpha }_{xx}$, and $\Gamma={\gamma }_{xx}^{r}+i{\gamma }_{xx}^{i}$.
The quantities $d_x$ and $d_{xx}$ are linear damping and amplitude dependent damping.  

The ellipsis in Eq.~(\ref{eq:fgrad}) indicates that in the presence of damping the operator $F_r$ must also include the temporal plane which undergoes synchrotron oscillations around an orbit which insures the recovery of energy due to the radiation process. In the code PTC, embedded in BMAD, the map can be computed with or without radiation and with coupling. To $7^{th}$ order we have computed the stable and unstable fixed points by solving the equation
\begin{eqnarray}
F_r\left({\vec{f}}\right)=0 \textrm{,}
\end{eqnarray}
using a Newton search which uses the result of low order perturbation theory as described in Sec.~\ref{theory}. 

It should be clear that all of this is automatically done in PTC using the analysis tools of FPP as described in Forest's book \cite{ef_book:2016}. Analytical formulas are not used and the map is computed using whatever complex model BMAD passes to the analysis tools via TPSA.

\section{Resonant driving terms}\label{appenRDT}
An individual resonance mode due to sextupoles is 
\begin{equation}
h_{i_1i_2i_3i_4i_5} (h_x^{+})^{i_1}(h_x^{-})^{i_2}(h_y^{+})^{i_3}(h_y^{-})^{i_4}\delta_e^{i_5}  \textrm{,}
\end{equation}
where $h_x^{\pm}=\sqrt{2J_x}e^{\pm i \phi_x}$ and $h_y^{\pm}=\sqrt{2J_y}e^{\pm i \phi_y}$ are the resonant basis, $\delta_e$ is the particle's energy offset, and $h_{i_1i_2i_3i_4i_5}$ is the coefficient. The resonance associated with each term is 
\begin{equation}
(i_1 - i_2) \nu_x +(i_3 - i_4) \nu_y = l \textrm{.}
\end{equation}
The sum of all resonance modes represents a normalized Harmiltonian near the zero closed orbit:
\begin{equation}
\sum h_{i_1i_2i_3i_4i_5} (h_x^{+})^{i_1}(h_x^{-})^{i_2}(h_y^{+})^{i_3}(h_y^{-})^{i_4}\delta_e^{i_5}+ c.c. \textrm{,} \label{eq:append} 
\end{equation}
where $i_1+i_2+i_3+i_4+i_5=n$ and $c.c.$ means complex conjugate. For a $3\nu_x$ resonance, comparing the terms including $J_x^{\frac{3}{2}}$ and $J_x^2$ in both Eq.~(\ref{eq:append}) and Eq.~(\ref{eq:h}) leads us to use $G=2h_{30000}$ and $\alpha_{xx}=-2h_{22000}$ in Eq.~(\ref{eq:action}). The minus sign is determined empirically from simulation. Here we list the formula of a resonance mode as a reminder and show the meaning of those indexing numbers in those terms. The detailed definition of the coefficient $h_{i_1i_2i_3i_4i_5}$ and its associated resonance mode can be found in Ref.~\citep{bengtsson:1997, wang:2012}. 

\section{Analytical formulas for Eq.~(\ref{eq:map})}\label{appenAna}
We have described above in some detail the lowest order Hamiltonian perturbation theory which in some cases can be carried out analytically. For example, all the objects in Eq.~(\ref{eq:map}) can be computed analytically under certain conditions. 

Assuming the system is linear except for sextupoles and no coupling, it is possible to derive analytically all the various operators in the absence of radiation. This was done partially by Forest in Section 5.2 in Ref.~\cite{ef_book:1997}. It is also possible by dimensional analysis involving the beta function to extract from this book some of the results in this appendix. The technique used here is the analytical technique of response functions described in that same book in Section 8.4. Unlike the usual Hamiltonian theory involving Fourier modes, everything here is ``exact'' within the model used and all the Fourier modes are included automatically. A code that checks these results and obtains machine precision agreement is available for the BMAD code. Here we list the derived analytic expressions for people who may be interested.

In this appendix, the formulas assume that $h =\sqrt{nJ_y}e^{- i\phi}$, where $n=2$ or $n=1$ and is related to the Poisson bracket $[h,\overline{h}]=-n\,i$.

In Eq.~(\ref{eq:map}), we have
\begin{eqnarray}
-2\pi {H}_{r}&=&-2\pi \delta {J}_{x}+{1 \over 2}{\alpha }_{xx}{J}_{x}^{2}+\Gamma {h}^{3}+\overline{\Gamma }{\overline{h}}^{3} \textrm{,}\label{eq:hrh}
\end{eqnarray}
and
\begin{align} 
{{\cal A}}_{s}&={{\cal A}}_{cs}\exp\left({:{F}_{21}{h}^{2}\overline{h}+{\overline{F}}_{21}h{\overline{h}}^{2}+{F}_{40}{h}^{4}}\right.\nonumber \\
&\left.{+\overline{{F}_{40}}{\overline{h}}^{4}+{F}_{31}^{0}{h}^{3}\overline{h}+\overline{{F}_{31}^{2}}h{\overline{h}}^{3}:}\right) \textrm{,}\label{eq:as}
\end{align}
where ${{\cal A}}_{cs}$ is the linear transformation in the Courant-Snyder format
\begin{align}
{{\cal A}}_{cs}=\left({\begin{array}{cc}\sqrt {\beta }&0\\
-{\alpha  \over \sqrt {\beta }}&{1 \over \sqrt {\beta }}\end{array}}\right) \textrm{.} \nonumber 
\end{align}
The amplitude dependent tune shift ${\alpha}_{xx} = {\alpha }_{1}+{\alpha }_{2}+{\alpha }_{3}$ has the following expressions,
\begin{align}
{\alpha}_{1} &= {1 \over 8}\left({\cot\left({{3\Delta  \over 2}}\right)-{3\Delta  \over 1-\cos\left({3\Delta }\right)}}\right) \\ 
&\times \SumInt\SumInt {\beta }_{{s}^{\prime }}^{3/2}{k}_{{s}^{\prime }}d{s}^{\prime }\ {\beta }_{s}^{3/2}{k}_{s}ds\ \cos\left({3\left({{\phi}_{{s}^{\prime }}-{\phi}_{s}}\right)}\right)\ \textrm{,} \nonumber \\
{\alpha }_{2} &=  {3 \over 8}\cot({\mu \over 2})  \\
 &\times\SumInt\SumInt {\beta }_{{s}^{\prime }}^{3/2}{k}_{{s}^{\prime }}d{s}^{\prime }{\beta }_{s}^{3/2}{k}_{s}ds\cos\left({{\phi}_{{s}^{\prime }}-{\phi}_{s}}\right)\ \textrm{,}  \nonumber  \\
{\alpha }_{3} &= {1 \over 4}\SumInt\SumInt {\beta }_{{s}^{\prime }}^{3/2}{k}_{{s}^{\prime}}d{s}^{\prime }{\beta }_{s}^{3/2}{k}_{s}ds\  \\
 &\times\left({3\sin\left({{\phi}_{{s}^{\prime }}-{\phi}_{s}}\right)+\sin\left({3\left({{\phi}_{{s}^{\prime }}-{\phi}_{s}}\right)}\right)}\right)\  \textrm{,}\nonumber  
\end{align}
where $\Delta=2\pi\delta$, $\mu=2\pi Q_{x}$, and, $k_s$, $\beta_s$ and $\phi_s$ are the sextupole strength, horizontal beta and phase functions at the location $s$, respectively. 
$\Gamma$ in Eq.~(\ref{eq:hrh}) has the following expression,
\begin{align} \Gamma &=
 {1 \over {n}^{3/2}\sqrt{8}}{i\Delta  \over e^{-i3\Delta }-1} \SumInt {\beta }_{s}^{3/2}{k}_{s}ds\ e^{-i3{\phi}_{s}}\ \textrm{,}
\end{align}
corresponding to $G$ in Eq.~(\ref{eq:h}). 

Other nonlinear terms such as $F_{21}$, $F_{40}$, and $F_{31}$ in Eq.~(\ref{eq:as}) have analytical formulas as well. They are necessary in order to find the canonical transformation ${\cal A}_{s}$. Here we list their analytic expressions.
\begin{align} 
F_{21} &=  {1 \over n^{3/2}\sqrt{8}\left({e^{-i\mu }-1}\right) } \SumInt {\beta}_{s}^{3/2} {k}_{s}ds e^{-i{\phi}_{s}} 
\end{align}
${F}_{40}^{}={F}_{40}^{1}+{F}_{40}^{2}$, where
\begin{align} 
{F}_{40}^{1}&= -{i \over {n}^{2}16}{1+{e}^{-i\mu } \over \left({1-{e}^{-i4\mu }}\right)\left({1-{e}^{-i\mu }}\right)} \nonumber \\
 & \times \SumInt\SumInt {\beta }_{{s}^{\prime }}^{3/2}{k}_{{s}^{\prime }}{\beta }_{s}^{3/2}{k}_{s}dsd{s}^{\prime }\ e^{-i ({\phi}_{{s}^{\prime }}+3{\phi}_{s})}\ \textrm{,} \\
{F}_{40}^{2}&=
-{i \over {n}^{2}32}{1+{e}^{-i\mu } \over \left({1-{e}^{-i4\mu }}\right)\left({1-{e}^{-i\mu }}\right)}  \nonumber \\
& \times \underbrace{\SumInt \SumInt}\limits_{s<{s}^{\prime }}^{} {\beta }_{{s}^{\prime }}^{3/2}{k}_{{s}^{\prime }}{\beta }_{s}^{3/2}{k}_{s}dsd{s}^{\prime } \nonumber \\
& \times\left\{{{e}^{-i({\phi }_{{s}^{\prime }}+3{\phi }_{s} )}-{e}^{-i ({\phi }_{s}+3{\phi }_{{s}^{\prime }})}}\right\} \textrm{.}
\end{align}
${F}_{31}={F}_{31}^{1}+{F}_{31}^{2}$, where

\begin{align} {F}_{31}^{1}&=
-{1 \over {n}^{2}8}i{1+{e}^{i\mu } \over \left({1-{e}^{-i2\mu }}\right)\left({1-{e}^{i\mu }}\right)}  \nonumber \\
 &\times \SumInt \SumInt {\beta }_{{s}^{\prime }}^{3/2}{k}_{{s}^{\prime }}{\beta }_{s}^{3/2}{k}_{s}dsd{s}^{\prime }\ e^{-i(-{\phi}_{{s}^{\prime }}+3{\phi}_{s})}\ \textrm{,} \\
{F}_{31}^{2}&=
-{i \over {n}^{2}16}{1+{e}^{i\mu } \over \left({1-{e}^{-i2\mu }}\right)\left({1-{e}^{i\mu }}\right)}\nonumber \\
 &\times \underbrace{\SumInt \SumInt}\limits_{s<{s}^{\prime }}^{} {\beta }_{{s}^{\prime }}^{3/2}{k}_{{s}^{\prime }}{\beta }_{s}^{3/2}{k}_{s}dsd{s}^{\prime }\nonumber \\
& \times\left\{{{e}^{-i(-{\phi }_{{s}^{\prime }}+3{\phi }_{s} )}-{e}^{-i (-{\phi }_{s}+3{\phi }_{{s}^{\prime }})}}\right\} \textrm{.}
 \end{align}

It is important to state that $H_r$ is not unique. Indeed, in a Fourier approach as performed in most text books \cite{sylee_book}, only a single term of the $3{\nu }_x$ resonance is left in the map. This is necessary for the co-moving Hamiltonian to be easily computed. But, as pointed out in Ref.~\cite{bazzani:1993}, it is only the interpolated Hamiltonian in the laboratory variables which is unique, i.e., what we call $K_r$ in Eq.~(\ref{eq:inter}). It is false to assume in a one-resonance computation that the normalized Hamiltonian $H_r$ is unique: this is only true in a non-resonant normal form because the gauge group is commutative. Thus neither the tune shift terms, nor the coefficients $\Gamma $ nor the fixed points computed by Eq.~(\ref{eq:angle}) are unique. Only the results in the laboratory coordinates are unique, i.e., the interpolated Hamiltonian $K_r$ or the vector field ${\Phi }_r$ (Eq.~(\ref{eq:interf})) which is the actual object computed by the library PTC even in the symplectic case.

\bibliography{tribs}

\end{document}